\documentclass[ACS,STIX1COL]{WileyNJD-v2}
\articletype{ORIGINAL ARTICLE}

\usepackage{mathtools}
\mathtoolsset{showonlyrefs=true}

\received{xxx}
\revised{xxx}
\accepted{xxx}

\newcommand{\onlinecite}[1]{\hspace{-1 ex} \nocite{#1}\citenum{#1}} % in case of non-revtex documentclass
\usepackage{bm}

\begin{document}

%\title{The Theory of Fluctuations: Dynamics of Classical and Quantum Systems
%}
\title{Classical and Quantum Theory of Fluctuations for Many-Particle Systems out of Equilibrium
}

\author[1]{E. Schroedter}
\author[1]{M. Bonitz}

\authormark{E. Schroedter  \textsc{et al}}

%\address[1]{\orgdiv{}\orgname{Lawrence Livermore National Laboratory},  \orgaddress{94550 \state{Livermore, CA}, \country{USA}}}

%\address[2]{\orgdiv{}\orgname{Center for Advanced Systems Understanding (CASUS)}, \orgaddress{D-02826 \state{G\"orlitz}, \country{Germany}}}
%\address[3]{\orgdiv{} \orgname{Helmholtz-Zentrum Dresden-Rossendorf}, \orgaddress{ D-01328\state{ Dresden}, \country{Germany}}}
% \affiliation{Center for Advanced Systems Understanding (CASUS), D-02826 G\"orlitz, Germany}
% \affiliation{Helmholtz-Zentrum Dresden-Rossendorf, D-01328 Dresden, Germany}

\address[1]{\orgdiv{Institut f\"ur Theoretische Physik und Astrophysik}, \orgname{Christian-Albrechts-Universit\"at zu Kiel}, \orgaddress{\state{Leibnizstra{\ss}e 15, 24098 Kiel}, \country{Germany}}}

%% \address[3]{\orgdiv{Org Division}, \orgname{Org Name}, \orgaddress{\state{State name}, \country{Country name}}}

\corres{*\email{schroedter@physik.uni-kiel.de}}
%% \presentaddress{This is sample for present address text this is sample for present address text}

\abstract{Correlated classical and quantum many-particle systems out of equilibrium are of high interest in many fields, including dense plasmas, correlated solids, and ultracold atoms. Accurate theoretical description of these systems is challenging both, conceptionally and with respect to computational resources.
%Warm dense matter--an exotic, highly compressed state on the boarder between solid and plasma phases is of high current interest, in particular for compact astrophysical objects, high pressure laboratory systems, and inertial confinement fusion. For many applications the interaction of quantum plasmas with energetic particles is crucial. Moreover, often the system is driven far out of equilibrium. In that case, there is high interest in time-dependent simulations to understand the physics, in particular, during thermalization. Recently a novel many-particle technique, the 
While for classical systems, in principle, exact simulations are possible via molecular dynamics, this is not the case for quantum systems. Alternatively, one can use many-particle approaches such as hydrodynamics, kinetic theory or nonequilibrium Green functions (NEGF). However, NEGF exhibit a very unfavorable cubic scaling of the CPU time with the number of time steps. An alternative is the G1--G2 scheme [N. Schlünzen et al., Phys. Rev. Lett.  \textbf{124}, 076601 (2020)] which allows for NEGF simulations with time linear scaling, however, at the cost of large memory consumption. The reason is the need to store the two-particle correlation function. This problem can be overcome for a number of approximations by reformulating the kinetic equations in terms of fluctuations -- an approach that was developed, for classical systems, by Yu.L. Klimontovich [JETP \textbf{33}, 982 (1957)]. Here we present an overview of his ideas and extend them to quantum systems. In particular, we demonstrate that this quantum fluctuations approach can reproduce the nonequilibrium $GW$ approximation [E. Schroedter \textit{et al.}, Cond. Matt. Phys. \textbf{25}, 23401 (2022)] promising high accuracy at low computational cost which arises from an effective semiclassical stochastic sampling procedure. We also demonstrate how to extend the approach to the two-time exchange-correlation functions and the density response properties. [E. Schroedter \textit{et al.}, Phys. Rev. B \textbf{108}, 205109 (2023)]. The results are equivalent to the Bethe--Salpeter equation for the two-time exchange-correlation function when the generalized Kadanoff-Baym ansatz with Hartree-Fock propagators is applied [E. Schroedter and M. Bonitz, submitted to phys. stat. sol. (b) 2024, arXiv:2312.15034].
}


\maketitle

%% \footnotetext{\textbf{Abbreviations:} ANA, anti-nuclear antibodies; APC, antigen-presenting cells; IRF, interferon regulatory factor}

\section{Introduction}\label{s:introduction}
Classical many-particle systems have been actively studied theoretically for more than one century with the main interest arising in dense gases, liquids and plasmas. The main methods have been kinetic theory, starting from Boltzmann's kinetic equation for the distribution function $f(r,p,t)$ and various generalizations such as the equilibrium and nonequilibrium hierarchy of reduced probability densities, $f_1, f_2, \dots$ (BBGKY-hierarchy). An independent direction of development was concerned with the dynamics of fluctuations, $\delta A=A-\overline{A}$, the deviations of random variables, $A$ from their mean value, $\overline{A}$. This classical fluctuations approach was pioneered in the Soviet Union by M.A. Leontovich, e.g. [\onlinecite{leontovich_jetp_35}]. The most sophisticated theory is due to  Yu.L.~Klimontovich, e.g.~[\onlinecite{klimontovich_jetp_57}]. Klimontovich was born 100 years ago, and, as we will show in this article, his theory even today provides an extensive sets of ``tools'' and ideas that can be applied to modern developments, for a historical overview see the article by Bonitz and Zagorodny in this volume~[\onlinecite{zagorodny_cpp_24}].

The main focus of the present paper is quantum many-body systems following an external excitation. This is a topic of high current interest in a variety of fields, which include dense plasmas, nuclear matter, ultracold atoms, and correlated solids. Multiple methods are available for simulating such systems, including real-time quantum Monte Carlo, density matrix renormalization group techniques, time-dependent density functional theory, and quantum kinetic theory. 
%In experimental investigations of many-particle systems, correlation functions of density or spin fluctuations and their corresponding dynamic structure factors play a central role, see e.g. Ref. \cite{giuliani_vignale_2005} for an overview. To accurately compute these quantities, taking correlation effects into consideration, various equilibrium simulations are employed. For correlated solids and warm dense matter, the most precise results are often obtained through quantum Monte Carlo simulations \cite{moreo_prb_93,lee_prb_03,assaad_prb_06,dornheim_physrep_18,dornheim_prl_18,dornheim_pop_23,bonitz_pop_20}, which also include the analysis of non-linear responses \cite{bonitz-etal.94pre,dornheim_prl_20,dornheim_prr_21}.
%
In addition to equilibrium simulations, a range of nonequilibrium approaches is available, such as nonequilibrium Green functions (NEGF), cf. [\onlinecite{keldysh64,stefanucci-book,balzer-book,bonitz_pss_19_keldysh}] and references therein,
dynamical mean field theory (DMFT), e.g., [\onlinecite{gull_prb_19,gull_prr_20}], and time-dependent density matrix renormalization group (DMRG), e.g., [\onlinecite{pereira_prb_12}].
Here, we focus on the NEGF approach, which offers a rigorous description of the quantum dynamics of correlated systems in multiple dimensions, e.g., [\onlinecite{schluenzen_prb16,schluenzen_jpcm_19}]. However, direct two-time NEGF simulations exhibit an unfavorable cubic scaling with simulation time $N_\mathrm{t}$ (number of time steps). Recently linear scaling with $N_\mathrm{t}$ has become feasible within the G1--G2 scheme [\onlinecite{schluenzen_prl_20,joost_prb_20}], which could be demonstrated even for advanced selfenergy approximations like $GW$ and the particle-paricle and particle-hole $T$-matrix approximations. Moreover, the full nonequilibrium version of the dynamically screened ladder approximation could be implemented, for lattice models [\onlinecite{joost_prb_22,donsa_prr_23}], for details of the scheme, see Ref.~\onlinecite{bonitz_pssb23}.
The advantage of the linear scaling in the G1--G2 scheme comes with a cost: the simultaneous propagation of the time-diagonal single-particle and correlated two-particle Green functions, $G_1(t)$ and $\mathcal{G}_2(t)$, demands a substantial computational effort for computing and storing all matrix elements of $\mathcal{G}_2$. For instance, the CPU time of $GW$-G1--G2 simulations scales with $N_\mathrm{b}^6$, where $N_\mathrm{b}$ denotes the size of the basis. 
%Moreover, going to a time-local formulation of the many-body problem severely restricts access to aforementioned correlation functions as they, for nonequilibrium systems, depend on multiple times. Thus, it is essential to explore alternative formulations that are more suitable for computational purposes, ideally without sacrificing accuracy.

For these reasons, it is of high interest to explore alternative concepts that provide the same accuracy of many-body simulations but at a significantly lower cost. Here the situation is similar to classical many-body systems for which many equivalent formulations of the nonequilibrium dynamics exist, as was discussed above.
In Ref. [\onlinecite{schroedter_cmp_22}], the present authors introduced an alternative formulation of the quantum many-body problem that is  based on a stochastic approach to the  dynamics of quantum fluctuations. Building upon earlier stochastic concepts in classical  kinetic theory by Klimontovich, that were mentioned above, e.g., [\onlinecite{klimontovich_jetp_57,klimontovich_jetp_72,klimontovich_1982}], and the work of Ayik, Lacroix~[\onlinecite{ayik_plb_08,lacroix_prb14,lacroix_epj_14}], and others, e.g., [\onlinecite{filinov_prb_2,polkovnikov_ap_10}], on stochastic approaches to describing the dynamics of quantum systems, an equation of motion for single-particle fluctuations, $\delta \hat{G}$, was derived that constitutes the basis of the quantum polarization approximation (QPA). It was shown that, in the weak coupling limit, the QPA is 
equivalent to the nonequilibrium $GW$ approximation of the G1--G2 scheme with additional exchange contributions.  

An advantage of the quantum fluctuations approach is that it allows for a straightforward extension from a time-local to a two-time description of the many-body system which is the basis for studying dynamic (frequency-dependent) response properties of correlated systems. Furthermore, in Ref.~\onlinecite{schroedter_23}, an extension of the stochastic approach to the so-called multiple ensembles (ME) approach was presented, which allows for the computation of commutators of operators and, thus, density response functions and their dynamic structure factors, both in the ground state and for systems far from equilibrium following an external excitation. Most importantly, this extension is applicable to large systems and long simulation times.
Finally, in Ref.~\cite{schroedter_23}
it was proven that the equivalence of the two-time quantum polarization approximation is equivalent to the Bethe--Salpeter equation of NEGF theory when the Hartree-Fock-GKBA is applied. This allows one to compute density and spin response properties and the dynamic structure factors  on the GW-Bethe-Salpeter level much more efficiently, using a stochastic implementation of the quantum polarization approximation.

%and the nonequilibrium $GW^\pm$ approximation extends from the time-local case to the two-time case. To this end, we consider the relation of the quantum fluctuations approach in its two-time formulation to the  and derive a two-time generalization of the correlated two-particle Green function, $\mathcal{G}_2(t,t')$, as well as its equations of motion. 

This paper is structured as follows. 
In Sec.~\ref{s:classical_fluctuations} we present an overview of Klimontovich's classical fluctuations approach focusing on the polarization approximation. There we demonstrate the strength of the approach by deriving the Balescu-Lenard kinetic equation (BLE) and discuss its quantum generalizations. The discussion of the  limitations of the BLE leads over to the derivation of a generalized quantum kinetic theory that is discussed in the remainder of the paper.
In Sec.~\ref{s:quantum_fluctuations}, we set up the necessary theoretical framework of our quantum fluctuations approach. We introduce nonequilibrium Green functions $G^{(s)}$ and the exchange-correlation functions $L^{(s)}$ and outline their relation to the quantum fluctuations approach. The central approximation is the quantum polarization approximation. In Sec.~\ref{s:stochastic_approach} we demonstrate how to practically evaluate the single-particle fluctuations equations using a stochastic approach and its dependence on the sampling probability. Next, in Sec.~\ref{s:numerics}, we present numerical illustrations of our quantum fluctuations approach by applying it to lattice models. Finally, we conclude with a summary and outlook in Sec.~\ref{s:discussion}.

\section{Classical fluctuations approach}\label{s:classical_fluctuations}
\subsection{Fluctuations, moments and distributions} \label{ss:classical_fluctuations:definitions}
We consider a nonrelativistic system of $N$ identical particles with mass $m$ and denote the phase space coordinate of the $i$-th particle with $x_i\coloneqq (\mathbf{r}_i,\mathbf{p}_i)$, where $\mathbf{r}_i$ and $\mathbf{p}_i$ denote its position and momentum, respectively. Further, the many-body system shall be described by a Hamiltonian of the form
\begin{equation}
    H\coloneqq \sum_{i=1}^N\left[ \frac{\mathbf{p}_i^2}{2m}+V_\mathrm{ext}(\mathbf{r}_i) \right]+\frac{1}{2}\sum_{\substack{i,j=1\\ i\neq j}}^N  W(\mathbf{r}_i,\mathbf{r}_j) \,,\label{eq:definition:classical_Hamiltonian}
\end{equation}
where $V_{\rm ext}$ denotes an external potential, e.g., an external electromagnetic field, and $W$ the pair interaction between two particles. \\
The microscopic state of the many-body system at time $t$ can be described using the so-called microscopic phase space density defined as
\begin{equation}
    N(x,t)\coloneqq \sum_{i=1}^N \delta[x-x_i(t)] \,,\label{eq:definition:microscopic_phase_space_density}
\end{equation}
with $x\coloneqq (\mathbf{r},\mathbf{p})$ and $x_i(t)$ denoting the trajectory of particle ``i''.\\
The quantity $N(x,t)$ was introduced by Klimontovich [\onlinecite{klimontovich_jetp_57}] and generalizes the concept of the charge density of a system of point charges, that is used in electromagnetism and field theory, to the phase space.
Using the microscopic phase space density, we can rewrite the Hamiltonian of the system, cf. Eq.~\eqref{eq:definition:classical_Hamiltonian}, as
\begin{equation}
    H=\int \left[\frac{\mathbf{p}^2}{2m}+V_\mathrm{ext}(\mathbf{r})\right]N(x,t)\,\mathrm{d}x+\frac{1}{2}\int W(\mathbf{r},\mathbf{r}') N(x,t)N(x',t)\,\mathrm{d}(x,x')-H_\mathrm{self}\,, \label{eq:classical_Hamiltonian_field_version}
\end{equation}
where $H_\mathrm{self}$ denotes the contribution due to the self-interaction which has to be subtracted to eliminate double countings of terms included in the second term on the r.h.s. In the following we neglect all contributions that arise due to the self-interaction. This representation of the classical Hamiltonian can be considered the analogue of the Hamiltonian (Hamilton operator) that is being used in quantum field theory, cf. Eq.~\eqref{eq:definition:Hamiltonian}. \\

As the exact microscopic state of the system is generally unknown it useful to introduce a probabilistic description where the points in $6N$-dimensional phase space are considered random
%\textcolor{red}{besser erklaeren! $N$ haengt von Anfangsbedingungen aller Trajektorien ab, die man im Prinzip angeben könnte. Das waere deterministische Dynamik, reiner Zustand. Eine andere Situation ist eine gemischte Gesamtheit mit gegebener Wahrsch-Verteilung der Mikrozustände.}
 variables %, i.e., $x_i(t)\rightarrow X_i(t)$ \textcolor{red}{Notation ist unklar, wann wird unten $X$ verwendet?} 
 that are associated with a probability density ($N$-particle distribution function) $P_N$ that is normalized to 1. Moreover, $P_N$ is assumed to be symmetric, i.e., 
\begin{equation}
    P_N(x_1,\dots, x_N,t)\equiv P_N(x_{\sigma(1)},\dots x_{\sigma(N)},t)\,,
\end{equation}
where $\sigma:\{1,\dots,N\}\rightarrow \{1,\dots,N\}$ denotes an arbitrary permutation.\\ 
Being a phase space function, $P_N$ obeys the Liouville equation where, on the r.h.s. we introduced the Poisson bracket\footnote{We define the Poisson bracket $\{\cdot,\cdot\}$ for two functions $f,g$ depending on $x_1,\dots,x_s$ as $\{f,g\}\coloneqq \sum_{i=1}^s[\nabla_{\mathbf{r}_i}f\cdot\nabla_{\mathbf{p}_i}g-\nabla_{\mathbf{p}_i}f\cdot\nabla_{\mathbf{r}_i}g ]$ }
\begin{equation}
    \partial_t P_N(x_1,\dots, x_N,t)=\left\{ H, P_N\right\}(x_1,\dots, x_N,t) \,.\label{eq:EOM:Liouville}
\end{equation}
In most cases, the full probability of an $N$-particle state is not needed, and it is sufficient to have information about lower dimensional states associated with a subset of $s$ particles. The corresponding $s$-particle distribution function is then defined as
\begin{equation}
    f_s(x_1,\dots,x_s,t)\coloneqq \frac{N!}{(N-s)!}\int P_N(x_1,\dots,x_s,x_{s+1},\dots, x_N,t)\,\mathrm{d}(x_{s+1},\dots,x_N)\,,\label{eq:definition:s-particle_distribution_function}
\end{equation}
where the prefactor describes the number of possibilities to choose $s$ particles out of $N$. Alternatively, the prefactor can be chosen to be $\mathcal{V}^s$, where $\mathcal{V}$ denotes the volume of the considered system. The latter definition turns out to be advantageous when considering macroscopic systems in the thermodynamic limit because then $s\ll N$. Moreover, we set $f\coloneqq f_1$. \\
Further, it is useful to to introduce correlation functions, $g_s$, for $s>1$, that are defined via the following relations (cluster expansion)
\begin{gather}
         f_2(x_1,x_2,t)\equiv f(x_1,t)f(x_2,t)+g_2(x_1,x_2,t)\,,\label{eq:definition:classical_two-particle_correlations}\\
     f_3(x_1,x_2,x_3,t)\equiv f(x_1,t)f(x_2,t)f(x_3,t)+f(x_1,t)g_2(x_2,x_3,t)+f(x_2,t)g_2(x_1,x_3,t)+f(x_3,t)g_2(x_1,x_2,t)+g_3(x_1,x_2,x_3,t)\,, \label{eq:definition:classical_three-particle_correlations}
\end{gather}
where the definitions for higher-order correlation functions follow analogously.  \\
Given an observable $A$ that depends on the phase space coordinates of all the particles as well as $x,t$,%, i.e., $A(X_1,\dots,X_N;x,t)\eqqcolon A(x,t)$,
its expectation value is given by
\begin{equation}
    \overline{A(x,t)}=\int A(x_1,\dots,x_N;x,t) P_N(x_1,\dots, x_N,t)\,\mathrm{d}(x_1\dots, x_N)\,. \label{eq:classical_expectation_value}
\end{equation}
This expression for the expectation value can be rewritten in terms of the $s$-particle distribution function given $A$ is an $s$-particle observable %, i.e., it can be expressed as
%\begin{equation}
%    A(X_1,\dots,X_N;x,t)=\sum_{1\leq i_1<\dots<i_s\leq N} A_s(X_{i_1},\dots, X_{i_s};x,t)
%\end{equation}
in the following way
\begin{equation}
    \overline{A(x,t)}=\frac{1}{s!}\int A_s(x_1,\dots,x_s;x,t) f_s(x_1,\dots,x_s,t)\,\mathrm{d}(x_1,\dots,x_s)\,.
\end{equation}
In most cases, it is sufficient to only consider the single- and two-particle distribution functions rather than $P_N$ to calculate the observables of interest such as the kinetic and interaction energy of a system. In addition to the expectation values of an observable $A$ important information is contained in the 
 fluctuation
\begin{equation}
    \delta A(x,t)\coloneqq A(x,t)-\overline{A(x,t)}\,,\label{eq:definition:fluctuations_classical_observables}
\end{equation}
which will be studied in detail below.

We now establish the connection of the reduced distribution functions and correlation functions to Klimontovich's
 microscopic phase space density, cf. Eq.~\eqref{eq:definition:microscopic_phase_space_density}. The first result is that the expectation of $N$ coincides with the single-particle distribution function: 
\begin{equation}
   \overline{N(x,t)} = f(x,t) \,. 
\end{equation}
The deviation of $N$ from the distribution function is given by the (classical) single-particle fluctuations, cf. Eq.~\eqref{eq:definition:fluctuations_classical_observables},
\begin{equation}
    \delta N(x,t)\coloneqq N(x,t)-\overline{N(x,t)} = N(x,t)-f(x,t)\,.\label{eq:definition:classical_single-particle_fluctuations}
\end{equation}
These fluctuations are the cornerstone of the classical theory of fluctuations as developed by Klimontovich. The next step of the theory is to consider products of fluctuations and their expectation values --  correlation functions of single-particle fluctuations:
\begin{equation}
    \Gamma_s(x_1,\dots, x_s,t)\coloneqq \overline{\delta N(x_1,t)\dots \delta N(x_s,t)}\,.\label{eq:definition:classical_s-particle_fluctuations}
\end{equation}
These quantities will be denoted as (classical) $s$-particle fluctuations. For the special cases of two- and three-particle fluctuations we will use separate notations, $\gamma$ and $\Gamma$, 
\begin{align}
    \gamma(x_1,x_2,t) \equiv \Gamma_2(x_1,x_2,t) &= \overline{ \delta N(x_1,t)\delta N(x_2,t)}\,,
    \label{eq:gamma-def}\\
    \Gamma(x_1,x_2,x_3,t) \equiv \Gamma_3(x_1,x_2,x_3,t) &= \overline{ \delta N(x_1,t)\delta N(x_2,t)\delta N(x_3,t)}
    \label{eq:gamma3-def}\,.
\end{align}
The $s$-particle fluctuations, $\Gamma_s$, are closely related to the correlation functions defined in Eqs.~\eqref{eq:definition:classical_two-particle_correlations} and \eqref{eq:definition:classical_three-particle_correlations}, as can be seen when considering the second and third moment of the microscopic phase space density and using Eq.~\eqref{eq:definition:classical_single-particle_fluctuations},
\begin{align}
\overline{N(x_1,t)N(x_2,t)} &=f(x_1,t)f(x_2,t)+\gamma(x_1,x_2,t)\,,\label{eq:classical_second_moment}\\
    \overline{N(x_1,t)N(x_2,t)N(x_3,t)} &= f(x_1,t)f(x_2,t)f(x_3,t)+f(x_1,t)\gamma(x_2,x_3,t)+f(x_2,t)\gamma(x_1,x_3,t)\\
    & \quad +f(x_3,t)\gamma(x_1,x_2,t)+\Gamma(x_1,x_2,x_3,t)\,, \label{eq:classical_third_moment}
\end{align}    
and we will establish the connections to $g_2$ and $g_3$ in the following.

\subsection{Properties of classical fluctuations} \label{ss:classical_fluctuations:properties}
As already mentioned, the first moment of the microscopic phase space density corresponds to the single-particle distribution function, i.e., $\overline{N(x,t)}=f(x,t)$. Similarly, the higher moments are connected to the higher-order distribution functions. For the second moment it holds
\begin{equation}
    \overline{N(x_1,t)N(x_2,t)}=f_2(x_1,x_2,t)+\delta(x_1-x_2)f(x_1,t)\,. \label{eq:classical_second_moment_two-particle_distribution_function}
\end{equation}
Thus, it follows for two-particle fluctuations, Eq.~\eqref{eq:gamma-def}, that they are related to two-particle correlations in the following way,
\begin{equation}
    \gamma(x_1,x_2,t)=\delta(x_1-x_2)f(x_1,t)+g_2(x_1,x_2,t)\,. \label{eq:classical_relation_gamma_g2}
\end{equation}
The first term on the r.h.s. of Eq.~\eqref{eq:classical_relation_gamma_g2} gives rise to so-called (classical) two-particle ``source fluctuations'' (using the term introduced by Klimontovich),
\begin{equation}
    \gamma^\mathrm{S}(x_1,x_2,t)\coloneqq \delta(x_1-x_2)f(x_1,t)\,,\label{eq:definition:classical_source_fluctuations}
\end{equation}
that are always present, even in an uncorrelated systems, where $g_s=0$, for all $s$. Hence, they can be interpreted as the source of two-particle fluctuations. Equation~\eqref{eq:classical_relation_gamma_g2} can then be rewritten as
\begin{equation}
    \gamma(x_1,x_2,t)=\gamma^\mathrm{S}(x_1,x_2,t)+g_2(x_1,x_2,t)\,. \label{eq:classical_relation_gamma_g2_2}
\end{equation}
Analogously, it is possible to find the relation of higher moments of the microscopic phase space density and higher-order distribution functions or, equivalently, the relation between $s$-particle fluctuations and correlation functions. For the third moment we have
\begin{gather}
    \overline{N(x_1,t)N(x_2,t)N(x_3,t)}=f_3(x_1,x_2,x_3,t)+\delta(x_1-x_2)f_2(x_2,x_3,t)+\delta(x_2-x_3)f_2(x_1,x_3,t)+\delta(x_1-x_3)f_2(x_1,x_2,t)\\+\delta(x_1-x_2)\delta(x_2-x_3)f(x_1,t)\,.
\end{gather}
Thus, the relation between three-particle fluctuations and the three-particle correlation function is given by
\begin{gather}
    \Gamma(x_1,x_2,x_3,t)\coloneqq \delta(x_1-x_2)g_2(x_2,x_3,t)+\delta(x_2-x_3)g_2(x_1,x_3,t)+\delta(x_1-x_3)g_2(x_1,x_2,t)\\+\delta(x_1-x_2)\delta(x_2-x_3)f(x_3,t)+g_3(x_1,x_2,x_3,t)\,.
\end{gather}
Here, we see that for an uncorrelated system, three-particle fluctuations also only depend on the single-particle distribution function. Analogous results hold for all higher-order quantities, but will not be explicitly stated here.\\
Another important property of fluctuations is their vanishing trace, i.e., it holds for all $s$-particle fluctuations that
\begin{equation}
    \int \Gamma_s(x_1,\dots,x_s,t)\,\mathrm{d}x_i=0 \,,\label{eq:classical_trace}
\end{equation}
for $i=1,\dots,s$, due to the conservation of the total particle number. 

\subsection{Dynamics of classical many-body systems in terms of $N$} \label{ss:classical_fluctuations:dynamics}
The equation of motion for the microscopic phase space density directly follows from Hamilton's equations and the conservation of the total particle number in phase space and is given by
\begin{equation}
    \left[ \partial_t+\mathbf{v}\cdot\nabla_\mathbf{r}+\mathbf{F}^\mathrm{M}(\mathbf{r},t)\cdot\nabla_\mathbf{p} \right]N(x,t)=0\,, \label{eq:EOM:N}
\end{equation}
where we introduced the microscopic force, $\mathbf{F}^\mathrm{M}\coloneqq -\nabla_\mathbf{r}[V+ U^\mathrm{M}]$, defined in terms of the gradient of the external potential $V$ and the microscopic mean-field potential $U^\mathrm{M}$ given by
\begin{equation}
   U^\mathrm{M}(\mathbf{r},t)\coloneqq \int W(\mathbf{r},\mathbf{r}')N(x',t)\,\mathrm{d}x'\,. \label{eq:definition_microscopic_mean-field_potential}
\end{equation}
Alternatively, it is possible to express Eq.~\eqref{eq:EOM:N} using the Poisson bracket and a microscopic Hamiltonian defined as
\begin{equation}
    H^\mathrm{M}(x,t)\coloneqq \frac{\mathbf{p}^2}{2m}+V(\mathbf{r})+U^\mathrm{M}(\mathbf{r},t)\,.
\end{equation}
allowing us to rewrite Eq.~(\ref{eq:EOM:N}) in form of a Liouville equation,
\begin{equation}
    \partial_t N(x,t)=\left\{H^\mathrm{M},N\right\}(x,t)\,.\label{eq:EOM:N_2}
\end{equation}
Taking the expectation value of Eq.~\eqref{eq:EOM:N_2} and using the bi-linearity of the Poisson bracket we find, for the single-particle distribution function, the following EOM 
\begin{equation}
    \partial_t f(x,t)=\left\{ H^\mathrm{MF},f_1\right\}(x,t)+\overline{\left\{\delta U^\mathrm{M},\delta N\right\}}(x,t)\,,\label{eq:EOM:single-particle_distribution_function}
\end{equation}
where $H^\mathrm{MF}\coloneqq \overline{H^\mathrm{M}}$ denotes the classical mean-field Hamiltonian. The last term on the r.h.s. of Eq.~\eqref{eq:EOM:single-particle_distribution_function} defines the so-called collision integral $I$ and depends on two-particle fluctuations, $\gamma$. It is explicitly given by
\begin{equation}
    I(x,t)\coloneqq \overline{\left\{\delta U^\mathrm{M},\delta N\right\}}(x,t)\equiv \int \nabla_\mathbf{r}W(\mathbf{r},\mathbf{r}')\cdot \nabla_\mathbf{p}\gamma(x,x',t)\,\mathrm{d}x'\,,\label{eq:definition:classical_collision_integral}
\end{equation}
and it describes the average of fluctuations interacting with the fluctuations mean-field $\delta U^{\rm M}$. \\
Moreover, the EOM for single-particle fluctuations directly follows from the difference of Eqs.~\eqref{eq:EOM:N_2} and \eqref{eq:EOM:single-particle_distribution_function} and the properties of the Poisson bracket, i.e., we find
\begin{equation}
    \partial_t\delta N(x,t)=\left\{ H^\mathrm{MF},\delta N\right\}(x,t)+\left\{\delta U^\mathrm{M},f\right\}(x,t)+\delta\left[\left\{\delta U^\mathrm{M},\delta N\right\}\right](x,t)\,,\label{eq:EOM:classical_single-particle_fluctuations}
\end{equation}
where the last term on the r.h.s. describes fluctuations of the interaction between fluctuations and the mean-field induced by fluctuations, i.e., it includes so-called second-order fluctuations defined as 
\begin{equation}
    \delta \gamma(x_1,x_2,t)\coloneqq \delta N(x_1,t)\delta N(x_2,t)-\gamma(x_1,x_2,t)\,.
\end{equation}
Using Eq.~\eqref{eq:EOM:classical_single-particle_fluctuations}, the EOMs for all fluctuations directly follow. For two-particle fluctuations it follows 
\begin{equation}
    \partial_t \gamma(x_1,x_2,t)=\left\{H^{(2),\mathrm{MF}},\gamma\right\}(x_1,x_2,t)+\pi(x_1,x_2,t)+C(x_1,x_2,t)\,, \label{eq:EOM:classical_two-particle_fluctuations}
\end{equation}
where we introduced the classical two-particle mean-field Hamiltonian defined as
\begin{equation}
    H^{(2),\mathrm{MF}}(x_1,x_2,t)\coloneqq H^\mathrm{MF}(x_1,t)+H^\mathrm{MF}(x_2,t)\,,
\end{equation}
and the (classical) polarization contribution given by
\begin{equation}
    \pi(x_1,x_2,t)\coloneqq \int \left\{ \gamma(x,x_2,t)\nabla_{\mathbf{r}_1} W(\mathbf{r},\mathbf{r}_1)\cdot\nabla_{\mathbf{p}_1}f(x_1,t)+\gamma(x,x_1,t)\nabla_{\mathbf{r}_2}W(\mathbf{r},\mathbf{r}_2)\cdot\nabla_{\mathbf{p}_2}f(x_2,t)\right\}\,\mathrm{d}x \,,\label{eq:definition:classical_polarization_term}
\end{equation}
and a term containing the coupling to three-particle fluctuations of the form 
\begin{equation}
    C(x_1,x_2,t)\coloneqq \int \left[ \nabla_{\mathbf{r}_1}W(\mathbf{r},\mathbf{r}_1)\cdot\nabla_{\mathbf{p}_1}+\nabla_{\mathbf{r}_1}W(\mathbf{r},\mathbf{r}_2)\cdot\nabla_{\mathbf{p}_2}  \right]\Gamma(x,x_1,x_2,t)\,\mathrm{d}x\,.
\label{eq:c-definition}
\end{equation}

\subsection{Approximations for classical fluctuations} \label{ss:classical_fluctuations:approximations}
\subsubsection{Approximations of moments} \label{sss:classical_fluctuations:approximation_of_moments}
The simplest type of approximations, within the hierarchy of classical fluctuations, is given by the ``approximations of moments''. Here, only fluctuations up to a certain order are considered, whereas higher-order contributions are neglected. Within the \textit{approximation of first moments} all contributions due to two-particle fluctuations are neglected which leads to the following EOM for the single-particle distribution function
\begin{equation}
    \partial_t f(x,t)=\left\{H^\mathrm{MF},f\right\}(x,t)\,,\label{eq:EOM:single-particle_distribution_function_1M}
\end{equation}
i.e., the collision term is neglected, and one recovers the standard (nonlinear) Vlasov equation. At the level of single-particle fluctuations this is equivalent to 
\begin{equation}
    \delta N(x,t)\equiv 0\,.
\end{equation}
Although fluctuations are always present, even for uncorrelated systems, cf. Eq~\eqref{eq:classical_relation_gamma_g2_2}, source fluctuations do not contribute to the collision integral, cf. Eq.~\eqref{eq:definition:classical_collision_integral}. Thus, fluctuations do not always need to significantly impact the single-particle dynamics of a system.\\

The next simplest approximation of this form is given by the \textit{approximation of second moments} where all contributions due to three-particle fluctuations are assumed to be vanishing, i.e., we have $C\approx 0$. The EOM for two-particle fluctuations, Eq.~\eqref{eq:EOM:classical_two-particle_fluctuations}, is then
\begin{equation}
    \partial_t \gamma(x_1,x_2,t)=\left\{H^{(2),\mathrm{MF}},\gamma\right\}(x_1,x_2,t)+\pi(x_1,x_2,t)\,. \label{eq:EOM:two-particle_fluctuations_2M}
\end{equation}
In the equation of single-particle fluctuations this is equivalent to neglecting all terms that are quadratic in $\delta N$, i.e., we have 
\begin{align}
    \partial_t \delta N(x,t)=\left\{ H^\mathrm{MF},\delta N\right\}(x,t)+\left\{\delta U^\mathrm{M},f\right\}(x,t)\,.
\end{align}
%\textcolor{red}{$U^{2M}$?}
While the approximations of moments have a clear mathematical definition, their physical relevance is limited. More important is the \textit{polarization approximation} as it allows to describe dynamical screening effects and leads to the Balescu-Lenard kinetic equation as we will show in Sec.~\ref{sss:ble}.

\subsubsection{Polarization approximation} \label{sss:classical_fluctuations:polarization_approximation}
In the polarization approximation (PA), three-particle fluctuations are not entirely neglected. Here, it is assumed that three-particle correlations are negligible and two-particle correlations are much smaller than two-particle fluctuations, i.e., $g_3\approx 0$ and $|g_2|\ll |\gamma|$ (weak coupling). This then leads to the following approximation for the three-particle fluctuations
\begin{equation}
    \Gamma(x_1,x_2,x_3,t)\approx \delta(x_1-x_2)\gamma(x_2,x_3,t)\,,
\end{equation}
which in turn leads to the following approximation for the three-particle coupling term, Eq.~\eqref{eq:c-definition},
\begin{equation}
    C(x_1,x_2,t) \approx R(x_1,x_2,t)\coloneqq  \delta(x_1-x_2) I(x_1,t)\,.
\end{equation}
It is now helpful to consider the EOM for two-particle source fluctuations, Eq.~\eqref{eq:definition:classical_source_fluctuations}, which is given by
\begin{equation}
    \partial_t \gamma^\mathrm{S}(x_1,x_2,t)= \left\{H^{(2),\mathrm{MF}},\gamma^\mathrm{S}\right\}(x_1,x_2,t)+R(x_1,x_2,t)\,.
\end{equation}
Hence, the EOM for the two-particle correlation function, within the polarization approximation, $g_2=\gamma-\gamma^\mathrm{S}$, is given by
\begin{equation}
    \partial_t g_2 (x_1,x_2,t)=\left\{H^{(2),\mathrm{MF}}, g_2 \right\}(x_1,x_2,t)+ \Psi(x_1,x_2,t)+\Tilde{\Pi}(x_1,x_2,t)\,, \label{eq:EOM_two-particle_correlation_function_PA}
\end{equation}
where $\Tilde{\Pi}$ denotes the (classical) correlations polarization contribution, which follows from $\pi$, cf. Eq.~\eqref{eq:definition:classical_polarization_term}, by the replacement $\gamma\rightarrow g_2$, whereas $\Psi$ denotes second-order scattering contributions and is defined as 
\begin{equation}
    \Psi(x_1,x_2,t)\coloneqq \nabla_{\mathbf{r}_1} W(\mathbf{r}_1,\mathbf{r}_2)\cdot\left[\nabla_{\mathbf{p}_1}+\nabla_{\mathbf{p}_2}\right]\left[ f(x_1,t)f(x_2,t) \right]\,. \label{eq:definition:classical_Psi}
\end{equation}
This contribution is an inhomogeneity that drives the build-up of correlations, i.e., given that a system is an uncorrelated initial state, $\Psi$ leads to non-vanishing the dynamics of $g_2$.\\

In the following, it is advantageous to represent the source fluctuations, $\gamma^S$, given by
Eq.~\eqref{eq:definition:classical_source_fluctuations} as a product of  single-particle source fluctuations, $\delta N^\mathrm{S}$, which are defined such that 
\begin{align}
    \overline{\delta N^\mathrm{S}(x,t)} &=0\,,\\
    \overline{\delta N(x,t)\delta N^\mathrm{S}(x',t)} &=0\,,\\
    \overline{\delta N^\mathrm{S}(x,t)\delta N^\mathrm{S}(x',t)} &= \gamma^\mathrm{S}(x,x',t) = \delta(x_1-x_2)f(x_1,t)\,.
\end{align}
Thus, by construction, $\delta N -\delta N^\mathrm{S}$ obeys an EOM that is equivalent to the EOM for $g _2$, cf. Eq.~\eqref{eq:EOM_two-particle_correlation_function_PA},
\begin{equation}
    \partial_t\left[ \delta N (x,t)-\delta N^\mathrm{S}(x,t) \right]= \left\{H^\mathrm{MF}, \delta N -\delta N^\mathrm{S}\right\}(x,t)+\left\{\delta U^\mathrm{M},f\right\}(x,t) \,, \label{eq:EOM:single-particle_fluctuations_PA}
\end{equation}
where $\delta N^\mathrm{S}$ obeys the homogeneous equation. 
%(\textcolor{red}{begruenden})
%\textcolor{red}{ das sollte man zeigen}

\subsection{Applications of the classical fluctuations approach. Balescu-Lenard equation} \label{ss:classical_fluctuations:applications}
The most important and impactful applications of Klimontovich's fluctuations approach were the derivation of a large variety of kinetic equations for gases and plasmas. As an example we consider his elegant derivation of the kinetic equation with a dynamically screened Coulomb potential which is commonly known as Balescu-Lenard equation [\onlinecite{balescu60,lenard60}]. This equation plays a fundamental role in the theory of collisional plasmas. It takes into account the long-range nature of the Coulomb interaction as well as the dynamical character (spectral or frequency dependence) of the field fluctuations and of collective plasma excitations which are described by a dynamic dielectric function, $\epsilon(q,\omega)$.

\subsubsection{Derivation of the Balescu-Lenard equation}\label{sss:ble}
We now consider a plasma of particles with charges $q_a$ interacting via the Coulomb interaction, i.e., $W\rightarrow V_{ab}$, where $V_{ab}(\mathbf{r},\mathbf{r}')\coloneqq q_aq_b/|\mathbf{r}-\mathbf{r}'|$ denotes the Coulomb potential. Consequently, the microscopic electric field follows from Eq.~\eqref{eq:definition_microscopic_mean-field_potential} and is given by 
\begin{equation}
    \mathbf{E}^\mathrm{M}(\mathbf{r},t) =\mathbf{E}_\mathrm{ext}(\mathbf{r},t)+ \sum_b q_b \int \frac{\mathbf{r}-\mathbf{r}'}{|\mathbf{r}-\mathbf{r}'|^3}\, N_b(x',t)\,\mathrm{d}x'\,,
\end{equation}
where $\mathbf{E}_\mathrm{ext}$ denotes the external electric field. Further, the collision integral, Eq.~\eqref{eq:definition:classical_collision_integral}, can then be expressed in terms of the fluctuations of the electric field, i.e., $\delta\mathbf{E}(\mathbf{r},t)\coloneqq \mathbf{E}^\mathrm{M}(\mathbf{r},t)- \overline{\mathbf{E}^\mathrm{M}(\mathbf{r},t)}$,
\begin{equation}
    I_a(x,t) = -\sum_b q_aq_b \int  \frac{\mathbf{r}-\mathbf{r}'}{|\mathbf{r}-\mathbf{r}'|^3} \nabla_\mathbf{p} \gamma_{ab}(x,x',t)\,\mathrm{d}x' = -q_a \nabla_\mathbf{p}\cdot\overline{\delta N_a(x,t)\delta\mathbf{E}(\mathbf{r},t)}\,.
\end{equation}
In order to capture dynamic (frequency-dependent) properties of the plasma it is necessary to consider the correlations of density and field fluctuations with a finite time delay. 
We, therefore, consider the more general case of different time arguments: $\overline{\delta N_a(x_1,t_1)\delta\mathbf{E}(\mathbf{r}_2,t_2)} \equiv \overline{\delta N_a\delta\mathbf{E}}(\mathbf{r}_1,\mathbf{p}_1,\mathbf{r}_2,t_1,t_2)$, and introduce relative and center-of-mass coordinates in space and time:
\begin{align}
    \Tilde{\mathbf{r}}&\coloneqq \mathbf{r}_1-\mathbf{r}_2\,, & \tau&\coloneqq t_1-t_2\,,\\
    \mathbf{R}&\coloneqq \frac{\mathbf{r}_1+\mathbf{r}_2}{2} \,, & T&\coloneqq\frac{t_1+t_2}{2}\,.
\end{align}
Thus, we can equivalently express any function depending on $\mathbf{r}_i$ and $\mathbf{t}_i$ in terms of these new coordinates,
\begin{equation}
  \overline{\delta N_a\delta\mathbf{E}}(\mathbf{r}_1,\mathbf{p}_1,t_1,\mathbf{r}_2,t_2) \rightarrow \overline{\delta N_a\delta\mathbf{E}}(\Tilde{\mathbf{r}},\mathbf{R},\mathbf{p}_1,\tau,T)\,.
\end{equation}
Additionally, we use the convention that, in the case of equal times, $t_1=t_2=t$, or equal positions, $\mathbf{r}_1=\mathbf{r}_2=\mathbf{r}$, we drop the relative coordinate.\\
By Fourier transforming the product of density and field fluctuations with respect to the relative position,
\begin{equation}
    \overline{\delta N_a \delta\mathbf{E}}(\Tilde{\mathbf{r}},\mathbf{R},\mathbf{p},\tau,T) = \int \overline{\delta N_a \delta\mathbf{E}}(\mathbf{k},\mathbf{R},\mathbf{p},\tau, T) e^{-\mathrm{i}\mathbf{k}\cdot \Tilde{\mathbf{r}}}\,\mathrm{d}\mathbf{k}\,,
\end{equation}
and, using that this product is a real function, i.e., we have $\overline{\delta N_a \delta\mathbf{E}}(\mathbf{k},\mathbf{R},\mathbf{p},\tau, T)= \big[\overline{\delta N_a \delta\mathbf{E}}(-\mathbf{k},\mathbf{R},\mathbf{p},\tau, T)\big]^*$, we get the following expression for the collision integral
\begin{equation}
    I_a(x,t) = -\frac{q_a}{(2\pi)^3}\nabla_\mathbf{p}\cdot\int \mathrm{Re}\left[ \overline{\delta N_a \delta\mathbf{E}}(\mathbf{k},\mathbf{r},\mathbf{p},t)\right]\,\mathrm{d}\mathbf{k}\,. \label{eq:collision_integral_E}
\end{equation}
In the following, we consider the polarization approximation where, the EOM for single-particle fluctuations, cf. Eq.~\eqref{eq:EOM:single-particle_fluctuations_PA}, takes the form
\begin{gather}
    \left(\partial_t + \mathbf{v}\cdot\nabla_\mathbf{r}+\mathbf{F}_a(\mathbf{r},t)\cdot \nabla_\mathbf{p}\right)\left[\delta N_a(x,t)-\delta N^\mathrm{S}_a(x,t)  \right]= - q_a \delta \mathbf{E}\cdot\nabla_\mathbf{p}f_a(x,t)\,, \label{eq:EOM:single-particle_fluctuations_PA(CP)}
\end{gather}
where we 
%set $f_a(x,t)\equiv f_{1,a}(x,t)$, for the single-particle distribution functions of different particle species and 
have, for the average force (Lorentz force),
\begin{equation}
    \mathbf{F}_a(\mathbf{r},t) = q_a \overline{\mathbf{E}^\mathrm{M}}(\mathbf{r},t)+\frac{q_a}{c}\mathbf{v}\times \mathbf{B}_\mathrm{ext}(\mathbf{r},t)\,,
\end{equation}
and $\mathbf{B}_\mathrm{ext}$ denotes the external magnetic field.
%which couples to Maxwell's equations
%\begin{equation}
%    \nabla \times \delta \mathbf{E}(\mathbf{r},t)=0\,,\quad \nabla\cdot \mathbf{E}(\mathbf{r},t) = 4\pi \sum_a q_a \int \delta N_a(x,t)\,\mathrm{d}\mathbf{p}\,.
%\end{equation}
Further, we consider two-time two-particle fluctuations defined as 
\begin{align}
    \gamma_{ab}(x_1,x_2,t_1,t_2) &\coloneqq \overline{\delta N_a(x_1,t_1)\delta N_b(x_2,t_2)}\,,\\
    \gamma_{ab}^\mathrm{S}(x_1,x_2,t_1,t_2) &\coloneqq \overline{\delta N^\mathrm{S}_a(x_1,t_1)\delta N^\mathrm{S}_a(x_2,t_2)}\,,
\end{align}
where the equal-time limit (initial condition) of the two-time source fluctuations is given by 
\begin{align}
\gamma^\mathrm{S}_{ab}(x_1,x_2,t_1,t_1)= \delta_{ab}\delta(x_1-x_2) f_a(x_1,t_1)\,. \label{eq:source-equal-time}   
\end{align}
 Two-particle source fluctuations then obey the following EOM
\begin{equation}
    \left[\partial_{t_1} +\mathbf{v}_1\cdot \nabla_{\mathbf{r}_1}+\mathbf{F}_a(\mathbf{r}_1,t_1)\cdot\nabla_{\mathbf{p}_1}\right]\gamma^\mathrm{S}_{ab}(x_1,x_2,t_1,t_2) =0 \,, \label{eq:EOM:two-time_source_fluctuations_PA(CP)}
\end{equation}
where $\gamma_{ab}^\mathrm{S}(t_1,t_2)$ evolves towards $t_1 > t_2$ (i.e. $\tau > 0$), starting from the initial condition \eqref{eq:source-equal-time}. This means, in the following, we consider a retarded function, $\gamma_{ab}^{S+}(\tau) \sim \theta(\tau)$.
Further, it is assumed that the  electric and magnetic fields are weak, so the contributions of the mean Lorentz force  to Eqs.~\eqref{eq:EOM:single-particle_fluctuations_PA(CP)} and \eqref{eq:EOM:two-time_source_fluctuations_PA(CP)} can be neglected, and we obtain
\begin{gather}
    \left(\partial_t + \mathbf{v}\cdot\nabla_\mathbf{r}\right)\left[\delta N_a(x,t)-\delta N^\mathrm{S}_a(x,t)  \right]= - q_a \delta \mathbf{E}\cdot\nabla_\mathbf{p}f_a(x,t)\,, \label{eq:EOM:single-particle_fluctuations_PA(CP)_2}\\
        \left(\partial_{t_1} +\mathbf{v}_1\cdot \nabla_{\mathbf{r}_1}\right)\gamma^\mathrm{S+}_{ab}(x_1,x_2,t_1,t_2) =0 \,,\quad t_1 > t_2\,. \label{eq:EOM:two-time_source_fluctuations_PA(CP)_2}
\end{gather}
Now we compute the spectral density of  the source fluctuations, i.e., the Fourier transform of $\gamma^\mathrm{S+}_{ab}$ with respect to the relative time and position. 
While the equal time limit of the source fluctuations is finite, cf. Eq.~\eqref{eq:source-equal-time}, we expect that the correlations of the fluctuations will decay to zero when $\tau = t_1-t_2 \to \infty$. (This excludes long lived and large scale correlations in the plasma that are e.g. due to bound states or turbulence and that require a separate discussion.)
The simplest way to achieve this is to
%to use a modified Fourier transform,
%\begin{align}
%    \gamma^S(\tau) \sim \int d \omega \,\gamma^S(\omega) \, e^{-i(\omega+i\delta)\tau}
%\end{align}
%we 
introduce a small dissipative correction, $\delta$, which obeys  $\omega_p \gg \delta>0$, into Eq.~\eqref{eq:EOM:two-time_source_fluctuations_PA(CP)_2}. This means that correlations in the plasma vanish for times larger than the correlation time [\onlinecite{bonitz_98teubner}] $\tau_{cor} \sim 2\pi/\omega_p$, and Eq.~\eqref{eq:EOM:two-time_source_fluctuations_PA(CP)_2} becomes 
\begin{gather}
%        \left(\partial_t + \mathbf{v}\cdot\nabla_\mathbf{r}\right)\left[\delta N_a(x,t)-\delta N^\mathrm{S}_a(x,t)  \right]= - q_a \delta \mathbf{E}\cdot\nabla_\mathbf{p}f_a(x,t)\,, \label{eq:EOM:single-particle_fluctuations_PA(CP)_2}\,,\\
        \left(\partial_{t_1} +\mathbf{v}_1\cdot \nabla_{\mathbf{r}_1}+\delta\right)\gamma^\mathrm{S+}_{ab}(x_1,x_2,t_1,t_2) =0 \,, \label{eq:EOM:two-time_source_fluctuations_PA(CP)_dissipative}
\end{gather}
where, in the final expressions, we will take the limit $\delta \rightarrow +0$.\\
The solution of Eq,~\eqref{eq:EOM:two-time_source_fluctuations_PA(CP)_dissipative} is given by
\begin{align}
    \gamma^\mathrm{S+}_{ab}(x_1,x_2,t_1,t_2) &= \delta_{ab} \delta[ \mathbf{r}_1-\mathbf{r}_2-\mathbf{v}_1(t_1-t_2)] \delta(\mathbf{p}_1-\mathbf{p}_2) e^{-\delta(t_1-t_2)}f_b(x_2,t_2)\,,
    \label{eq:solution_source_fluctuations}\\
   \rightarrow \gamma^\mathrm{S+}_{ab}(\Tilde{\mathbf{r}},\mathbf{R},\mathbf{p}_1,\mathbf{p}_2,\tau,T) &=  \delta_{ab}\delta(\Tilde{r}-\mathbf{v}_1\tau) \delta(\mathbf{p}_1-\mathbf{p}_2)e^{-\delta \tau} f_b(\mathbf{R}-\Tilde{\mathbf{r}}/2,\mathbf{p}_2,T-\tau/2)\,.
\end{align}
Analogously, we can treat the advanced function $\gamma^{\mathrm{S}-}(x_1,x_2,t_1,t_2)\sim\theta(-\tau)$, which obeys the following EOM 
\begin{equation}
            \left(\partial_{t_2} +\mathbf{v}_2\cdot \nabla_{\mathbf{r}_2}+\delta\right)\gamma^\mathrm{S-}_{ab}(x_1,x_2,t_1,t_2) =0 \,,\quad t_2> t_1\,, \label{eq:EOM:two-time_source_fluctuations_PA(CP)_dissipative_2}
\end{equation}
and thus has a solution given by
\begin{align}
    \gamma^\mathrm{S-}_{ab}(x_1,x_2,t_1,t_2) &= \delta_{ab} \delta[ \mathbf{r}_1-\mathbf{r}_2-\mathbf{v}_1(t_1-t_2)] \delta(\mathbf{p}_1-\mathbf{p}_2) e^{-\delta(t_2-t_1)}f_b(x_1,t_1)\,,
    \label{eq:solution_source_fluctuations}\\
   \rightarrow \gamma^\mathrm{S-}_{ab}(\Tilde{\mathbf{r}},\mathbf{R},\mathbf{p}_1,\mathbf{p}_2,\tau,T) &=  \delta_{ab}\delta(\Tilde{r}-\mathbf{v}_1\tau) \delta(\mathbf{p}_1-\mathbf{p}_2)e^{\delta \tau} f_b(\mathbf{R}+\Tilde{\mathbf{r}}/2,\mathbf{p}_1,T+\tau/2)\,.
\end{align}
Two-time source fluctuations are then given by a combination of the retarded and advanced function, i.e., 
\begin{equation}
    \gamma_{ab}^\mathrm{S}(x_1,x_2,t_1,t_2) = \begin{cases}
        \gamma^\mathrm{S+}(x_1,x_2,t_1,t_2)\,, & t_1>t_2\,,\\
        \delta_{ab}\delta(x_1-x_2) f_a(x_1,t_1) \,, & t_1=t_2\,, \\
        \gamma_{ab}^\mathrm{S-}(x_1,x_2,t_1,t_2)\,, & t_1<t_2\,.
    \end{cases}
\end{equation}
In the following, we assume that the fluctuations are stationary and uniform in space, i.e., there is no dependence on the center-of-mass coordinate $\mathbf{R}$ and on the macroscopic time $T$. Then, source fluctuations are of the form
\begin{equation}
    \gamma^\mathrm{S}_{ab}(\Tilde{\mathbf{r}}, \mathbf{p}_1,\mathbf{p}_2,\tau) = \delta_{ab} \delta[\Tilde{\mathbf{r}}-\mathbf{v}_1\tau] \delta(\mathbf{p}_1-\mathbf{p}_2) e^{-\delta|\tau|}f_a(\mathbf{p}_1)\,,
\end{equation}
and their Fourier transform is given by 
%\begin{align}
 %   \gamma^{\mathrm{S},+}_{ab}(\mathbf{k},\mathbf{p}_1,\mathbf{p}_2,\omega)\coloneqq \int_0^\infty \int \gamma^\mathrm{S+ }_{ab}(\Tilde{\mathbf{r}},%\mathbf{p}_1,\mathbf{p}_2,\tau) e^{\mathrm{i}(\omega\tau-\mathbf{k}\cdot\Tilde{\mathbf{r}})}\,\mathbf{d}\mathbf{k}\mathrm{d}\tau= \delta_{ab}%\delta(\mathbf{p}_1-\mathbf{p}_2) \frac{\mathrm{i}}{\omega- \mathbf{k}\cdot \mathbf{v}+\mathrm{i}\delta}f_a(\mathbf{p}_1)\,.
%\end{align}
%Using that, for a stationary process, \textcolor{red}{das ist unklar: warum wird im Folgenden diese Größe benutzt? Ich vermute, das hier retardierte und %avancierte Funktion addiert werden, so dass die Funktion fuer alle tau gilt}
\begin{equation}
    \gamma_{ab}^\mathrm{S}(\mathbf{k},\mathbf{p}_1,\mathbf{p}_2,\omega) = \delta_{ab}\delta(\mathbf{p}_1-\mathbf{p}_2) \frac{2\delta}{(\omega-\mathbf{k}\cdot\mathbf{v}_1)^2+\delta^2}\,f_a(\mathbf{p}_1)\,.
\end{equation}
In the limit $\delta \rightarrow +0$, the  spectral density becomes
\begin{equation}\label{eq:gammas-result}
    \gamma_{ab}^\mathrm{S}(\mathbf{k},\mathbf{p}_1,\mathbf{p}_2,\omega)=2\pi\delta_{ab} \delta(\mathbf{p}_1-\mathbf{p}_2) \delta(\omega- \mathbf{k}\cdot\mathbf{v}_1) f_a(\mathbf{p}_1)\,,
\end{equation}
and exhibits the familiar resonance behavior that couples the particle velocity with the phase velocity $\omega/k$ of an excitation.

We now apply this result to obtain the spectral functions of particle-field and field fluctuations. Consider first the Fourier transform of the (source) fluctuations of the electric field: 
\begin{eqnarray}\label{eq:delta-es}
    \delta \mathbf{E}(\mathbf{k},\omega) &=& -\frac{4\pi \mathrm{i} \mathbf{k}}{k^2}\sum_a q_a \int \delta N_a(\mathbf{k},\mathbf{p},\omega) \,\mathrm{d}\mathbf{p}\,,\\
    \delta \mathbf{E}^{\mathrm{S}}(\mathbf{k},\omega) &=& -\frac{4\pi \mathrm{i} \mathbf{k}}{k^2}\sum_a q_a \int \delta N^{\mathrm{S}}_a(\mathbf{k},\mathbf{p},\omega) \,\mathrm{d}\mathbf{p}\,,\label{eq:delta-e}    
\end{eqnarray}
which, upon using Eq.~\eqref{eq:gammas-result}, yields the following expressions
\begin{gather}
    \overline{\delta N^\mathrm{S}_a\delta\mathbf{E}^\mathrm{S}}(\mathbf{k},\mathbf{p},\omega) = \frac{8\pi^2 \mathrm{i}q_a\mathbf{k}}{k^2} \delta(\omega- \mathbf{k}\cdot \mathbf{v}) f_a(\mathbf{p})\,,\label{eq:source_density/electric_field_LB}\\
    \overline{\delta \mathbf{E}^\mathrm{S}\cdot\delta\mathbf{E}^\mathrm{S}}(\mathbf{k},\omega) = \sum_a \left(\frac{4\pi q_a}{k}\right)^2 \int \delta(\omega-\mathbf{k}\cdot \mathbf{v}) f_a(\mathbf{p})\,\mathrm{d}\mathbf{p}\,. \label{eq:source_electric_field_LB}
\end{gather}
Next, we consider Eq.~\eqref{eq:EOM:single-particle_fluctuations_PA(CP)_2} and introduce a dissipative term, similar to our previous considerations, i.e., 
\begin{equation}
     \left(\partial_t + \mathbf{v}\cdot\nabla_\mathbf{r}+\delta\right)\left[\delta N_a(x,t)-\delta N^\mathrm{S}_a(x,t)  \right]= - q_a \delta \mathbf{E}\cdot\nabla_\mathbf{p}f_a(\mathbf{p})\,. \label{eq:EOM:single-particle_fluctuations_PA(CP)_dissipative}
\end{equation}
For the computation of the particle-field correlations which proceed on the time scale $\tau_{cor}$, we may assume that 
$f_a$ is time independent because it changes significantly only on the scale of the relaxation time where $t_{rel}\gg \tau_{cor}$. Similarly, $f_a$ may be considered nearly uniform on spatial scales of the plasmas oscillations (wavelength).
Then, the Fourier transform of the solution of Eq.~\eqref{eq:EOM:single-particle_fluctuations_PA(CP)_dissipative} is easily computed:
\begin{equation}
    \delta N_a(\mathbf{k},\mathbf{p},\omega) = \delta N^\mathrm{S}(\mathbf{k},\mathbf{p},\omega) -\frac{\mathrm{i}q_a}{\omega- \mathbf{k}\cdot \mathbf{v}+i\delta} \delta\mathbf{E}(\mathbf{k},\omega)\cdot\nabla_\mathbf{p}f_a(\mathbf{p}) \,. \label{eq:solution:density_fluctuations}
\end{equation}
%Combining Eq.~\eqref{eq:source_electric_field_LB} and 
We now eliminate from
Eq.~\eqref{eq:solution:density_fluctuations}
the fluctuations $\delta N_a$ and $\delta N_a^\mathrm{S}$ in favor of the corresponding field fluctuations. To this end we integrate Eq.~\eqref{eq:solution:density_fluctuations}
over the momentum and use Eqs.~\eqref{eq:delta-e} and \eqref{eq:delta-es}, which yields a linear relation between the fluctuations of the electric field and its source fluctuations
\begin{equation}
    \epsilon(\mathbf{k},\omega) \delta \mathbf{E}(\mathbf{k},\omega) = \delta \mathbf{E}^\mathrm{S}(\mathbf{k},\omega)\,, \label{eq:dE^s-dE}
\end{equation}
with the proportionality coefficient $\epsilon$ which is nothing but the  (retarded) dielectric function familiar from electrodynamics and plasma theory. Its result is given by 
\begin{gather}
    \epsilon(\mathbf{k},\omega) \coloneqq 1+\sum_{a} V_{aa}(k) \Pi_{a}(\mathbf{k},\omega)\equiv  1+ V(k) \Pi(\mathbf{k},\omega)\,,\label{eq:definiton_dielectric_function}\\
    \Pi_{a}(\mathbf{k},\omega ) \coloneqq  \int\frac{\mathbf{k}\cdot \nabla_\mathbf{p}f_a(\mathbf{p})}{\omega- \mathbf{k}\cdot\mathbf{v}+\mathrm{i}\delta}\,\mathrm{d}\mathbf{p}\,,\label{eq:vlasov-p}
\end{gather}
where $V_{ab}(k)= 4\pi q_aq_b/k^2$ is the Fourier transform of the Coulomb potential and $\Pi_{a}(\mathbf{k},\omega)$ denotes the (retarded) polarization function of an ideal plasma (which implies that $\delta \rightarrow +0$). Obviously, with Eqs.~\eqref{eq:vlasov-p} and \eqref{eq:definiton_dielectric_function} we have recovered the familiar Vlasov polarization and longitudinal dielectric function of a classical ideal plasma.

With these results for the spectral function of the fluctuations we now return to the kinetic equation and its collision integral, Eq.~\eqref{eq:collision_integral_E}.
Combining our results, i.e., Eqs.~\eqref{eq:source_density/electric_field_LB}, \eqref{eq:source_electric_field_LB}, \eqref{eq:solution:density_fluctuations} as well as \eqref{eq:dE^s-dE}, and using that $\delta\mathbf{E}$ and $\mathbf{k}$ are parallel, we find
%\begin{equation}
%    \mathrm{Re}\left[ \overline{\delta N\delta \mathbf{E}}(\mathbf{k},\mathbf{p},\omega)\right] = -\pi q_a\delta(\omega- \mathbf{k}\cdot \mathbf{v})\frac{\mathbf{k}}{k^2} \overline{\delta\mathbf{E}\cdot\delta\mathbf{E}}(\mathbf{k},\omega) \mathbf{k}\cdot \nabla_{\mathbf{p}}f_a(\mathbf{p})- \frac{8\pi^2 q_a}{k^2}\delta(\omega-\mathbf{k}\cdot v) \frac{\mathrm{Im}\left[\epsilon(\mathbf{k},\omega)\right]}{|\epsilon(\mathbf{k},\omega)|^2}f_a(\mathbf{p}) 
%\end{equation}
\begin{align}
      \mathrm{Re}\left[ \overline{\delta N\delta \mathbf{E}}(\mathbf{k},\mathbf{p})\right] = -\frac{16\pi^3q_a}{k^4} \mathbf{k}\sum_b q_b^2 \int \frac{\delta[\mathbf{k}\cdot (\mathbf{v}- \mathbf{v}')]}{|\epsilon(\mathbf{k},\mathbf{k}\cdot\mathbf{v})|^2} \mathbf{k}\cdot \left[\nabla_\mathbf{p}f_a(\mathbf{p})f_b(\mathbf{p}')-\nabla_{\mathbf{p}'}f_b(\mathbf{p}')f_a(\mathbf{p})\right]\,\mathrm{d}\mathbf{p}'\,,
      \label{eq:result-dnde}
\end{align}
where we integrated over $\omega$, i.e., we set $\mathrm{Re}[ \overline{\delta N\delta \mathbf{E}}(\mathbf{k},\mathbf{p})]\equiv \int  \mathrm{Re}[ \overline{\delta N\delta \mathbf{E}}(\mathbf{k},\mathbf{p},\omega)]\,\mathrm{d}\omega$. Since the kinetic equation describes the evolution of the single-particle distribution function which proceeds on times much larger than the correlation time and lengths larger than the wavelength, we may restore time and space dependence in the single-particle distribution functions and the collision integral. Inserting the result \eqref{eq:result-dnde} into Eq.~\eqref{eq:collision_integral_E} we find the following result for the collision integral, 
\begin{equation}\label{eq:lb-col-int}
    I_a(x,t) = \frac{1}{8\pi^2}\sum_b \int \int \left[V^s_{ab}(k,\mathbf{k}\cdot\mathbf{v})\right]^2\delta[\mathbf{k}\cdot (\mathbf{v}- \mathbf{v}')]\cdot (\mathbf{k}\cdot\nabla_\mathbf{p})  \mathbf{k}\cdot \left[\nabla_\mathbf{p}f_a(\mathbf{p})f_b(\mathbf{p}')-\nabla_{\mathbf{p}'}f_b(\mathbf{p}')f_a(\mathbf{p})\right] \,\mathrm{d}\mathbf{p}'\mathrm{d}\mathbf{k}\,.
\end{equation}
%\textcolor{red}{hier die Ableitung kompakt, nur die wichtigsten Schritte und Gleichungen}\\
%The final result for the collision integral is (we omit the time dependence of the distribution functions)
%\begin{align}
%    I_a(\textbf{r},\textbf{p}) &= 2 \sum_b n_b \frac{\partial}{\partial p_i}\int d\textbf{q}d\textbf{p}' \frac{q_i q_j V^2_{ab}(q)}{|\epsilon(\textbf{q}\cdot \textbf{v},q)|^2}
%    \,\delta(\textbf{q}\textbf{v}-\textbf{q}\textbf{v}')\left\{ \frac{\partial f_a(\textbf{r},\textbf{p})}{\partial p_j}f_b(\textbf{r},\textbf{p}') - 
%    \frac{\partial f_b(\textbf{r},\textbf{p}')}{\partial p'_j}f_a(\textbf{r},\textbf{p})
%    \right\}\,,\label{eq:ilb}\\
%    \epsilon(q,\omega) &= 1 - V(q) \Pi(q,\omega)\,,\quad V_{ab}(q) = \frac{4\pi e_ae_b}{q^2}\,,\quad \Pi(q,\omega) = - \sum_a n_a \int d \textbf{p} \:\frac{\textbf{q} \cdot\frac{\partial f_a}{\partial \textbf{p}}}{\omega - \textbf{q}\cdot \textbf{v} + i\delta}\,, \label{eq:vlasov-df}
%\end{align}
where the delta function reflects kinetic energy conservation and the distribution functions are to be understood as nonequilibrium functions, $f_a(t)$.
This is the collision integral that was first derived by Lenard and Balescu [\onlinecite{lenard60,balescu60}] and describes charged particle scattering in a plasma medium. Scattering is treated perturbatively and involves the square of the pair potential (second Born approximation). However, screening and collective effects (including plasmons and instabilities) in the plasma lead to a replacement of the Coulomb potential $V(k)$ by the dynamically screened Coulomb potential $V^s(k,\omega)$,
\begin{align}
    V^s_{ab}(k,\omega) &= \frac{V_{ab}(k)}{|\epsilon(k,\omega)|}\,,\label{eq:vs-definition}\\
    \delta \textbf{E}(k,\omega) &= \frac{\delta\textbf{E}^\mathrm{S}(k,\omega)}{\epsilon(k,\omega)}\,,\label{eq:de-ds}
\end{align}
Simpler versions of collision integrals in plasmas or condensed matter, such as the Landau collision integral, follow as a special case from Eq.~\eqref{eq:lb-col-int}. In fact, use of the bare Coulomb potential, $V(k)$, in the collision integral leads to divergent results -- the $k$-integration diverges logarithmically at small and large $k$ giving rise to the ``Coulomb logarithm'', $\Lambda=\ln(k_{\rm max}/k_{\rm min})$. The small-$k$ divergency is due to the long range of the interaction, and this problem is ``fixed'' in the Landau equation by replacing $V(k)$ by a phenomenologically statically screened potential. With the Balescu-Lenard result \eqref{eq:lb-col-int}, no phenomenological corrections are required and screening appears automatically. The Landau equation then follows as the static long wavelength limit of the Vlasov dielectric function \eqref{eq:definiton_dielectric_function}, $\lim_{q\to 0}\epsilon(q,\omega=0)$, giving rise to a Debye-screened Coulomb potential,  so that $k_{\rm min} \to 1/r_D$.
At the same time, the divergence of the $k$-integration at large $k$ remains also in the case of the collision integral \eqref{eq:lb-col-int} and is removed by a phenomenological cutoff, $k_{\rm max}$. This problem is naturally ``fixed'' by taking into account quantum effects that appear at small distances, as we show in Sec.~\ref{sss:ble-quantum}.

Furthermore, with Eq.~\eqref{eq:de-ds} we reformulated our previous result, Eq.~\eqref{eq:dE^s-dE} in a physically more intuitive relation: collective plasma effects that are condensed in the dielectric function lead to a replacement of the electric source field fluctuations (which are generated by $\delta N^\mathrm{S}$ and are always present in the system) by the fluctuations $\delta \textbf{E}$. Finally, relations ~\eqref{eq:vs-definition} and \eqref{eq:de-ds} are rather general and not restricted to the case of the Vlasov dielectric function. Deriving improved results that take into account quantum effects, cf. Sec.~\ref{sss:ble-quantum} or correlations allows for further improvement of kinetic theory. For example, strong coupling effects (large angle and multiple scattering) can be treated within the dynamically screened ladder approximation, e.g. [\onlinecite{joost_prb_22}] or with quantum Monte Carlo methods, e.g. [\onlinecite{dornheim_prl_18}], see also Sec.~\ref{s:discussion}.

\subsubsection{Extension of the Balescu-Lenard equation to quantum systems}\label{sss:ble-quantum}
The quantum analogue of the Balescu-Lenard equation has been broadly studied in semiconductor optics where this equation has been known as ``quantum Boltzmann equation'' that has been derived by many authors, see e.g. the text book of Kadanoff and Baym~[\onlinecite{kadanoff-baym}]. This equation for the single-particle distributions of electrons and holes contains the collision integral ($a, b=e,h$)
\begin{align}
    I_a(k_1) &= \Gamma^a_{in}(k_1)f_a(k_1) - \Gamma^a_{out}(k_1)f_a^>(k_1)\,,\quad f_a^>(k)\equiv 1-f_a(k)\,,\label{eq:i-qbl}\\
    \Gamma^b_{in}(k_1) &= \frac{4\pi}{\hbar} \sum_{a,k_2,k_3,k_4} \left|V_{ab}^s\left[k_2-k_1,E^b(k_2)-E^b(k_1)\right]\right|^2f_b(k_2)f^>_\alpha(k_3)f_\alpha(k_4)\delta_{k_1+k_3,k_2+k_4}\delta\left[E^a(k_1)-E^b(k_2)+E^a(k_3)-E^b(k_4)\right]\,,\label{eq:gamma-qlb}\\
    V_{ab}^s(q,\omega) &= \frac{V_{ab}(q)}{|\epsilon(q,\omega)|}\,,\quad \epsilon(q,\omega) = 1 - V(q)\Pi(q,\omega)\,,\quad V_{ab}(q) = \frac{4\pi e_ae_b}{q^2}\,,\quad \Pi(q,\omega) = 2\sum_{\alpha k}\frac{f_\alpha(k)-f_\alpha(|\textbf{q}+\textbf{k}|)}{E^\alpha(k)-E^\alpha(|\textbf{q}+\textbf{k}|)+\hbar\omega + i\delta}\,,
\end{align}
where $\Gamma_{in,out}$ are scattering rates, and $\Gamma_{out}$ follows from $\Gamma_{in}$ by exchanging $f \leftrightarrow f^>$. Carrier scattering is treated in perturbation theory (second Born approximation, SOA) where the Coulomb potential with the Fourier transform $V(q)$ is screened by the dielectric function $\epsilon(q,\omega)$ exactly like in the classical case, cf. Eq.~\eqref{eq:lb-col-int}. The main difference are quantum effects. They lead to the appearance of Pauli blocking in the scattering rates (the factors $f^>$) which prevent the divergence of the integral at large momenta (in the present notation, this corresponds to large differences $k_1-k_2$). The second main difference is in the longitudinal retarded polarization ($\delta\to +0$) which is now that of an ideal Fermi gas $\Pi$ (Lindhard or RPA polarization). It is easily verified that the classical (long wavelength) limits of the collision integral and of the polarization coincide with the classical expressions \eqref{eq:lb-col-int}
and \eqref{eq:vlasov-p}, respectively.

\subsubsection{Extension of the Balescu-Lenard equation to short time scales. Dynamics of correlations}\label{sss:ble-short}

The coupled quantum kinetic equations for electrons and holes with the collision integral \eqref{eq:i-qbl}
were solved numerically by Binder \textit{et al.} [\onlinecite{scott_prl_92,binder_prb_92}]. They observed that the dynamically screened Coulomb potential may give rise to extremely high scattering rates, $\Gamma$, and consequently to surprisingly fast (within a few femtoseconds) dephasing and thermalization of optically excited semiconductors. This was explained by plasmon undamping resulting from zeroes of the dielectric function which was further analzyed in Ref.~[\onlinecite{scott-etal.94prb}]. The key is that, in the collision integral and also in the dielectric function, the nonequilibrium distribution functions appear which give rise to additional weakly damped plasmons -- zeroes of Re$\epsilon(q,\omega)$ -- which act as additional scattering channels.
Indeed, an accelerated relaxation, in case of undamped nonequilibrium plasmons, was confirmed experimentally in Refs.~[\onlinecite{lampin_prb_99,bonitz_prb_0}]. However, the experimental time scales of the dephasing and relaxation in semiconductors turned out to be much longer than predicted in Ref.~\onlinecite{scott_prl_92} which stimulated investigations of the validity range of the quantum Balescu-Lenard equation and similar quantum kinetic equations. 

The analysis revealed that, in case of rapid dynamics of the distribution function $f_a(k,t)$, the use of the standard RPA (or Vlasov) dielectric function with the current distribution function, $\epsilon[q,\omega;f(t)]$ violates the time scale separation, $\tau_{cor}\ll t_{rel}$. If this inequality does not hold, the dielectric function has to be generalized such that it includes correlation dynamics and time-dependent build up of dynamical screening. This has led to generalized quantum kinetic equations that include non-Markovian (time retardation or memory) effects, e.g. ~[\onlinecite{bonitz_98teubner,bonitz-etal.96jpcm}]. This was achieved by the development of nonequilibrium Green functions (NEGF) and reduced density operator methods. One result was the theoretical simulation of the buildup of screening by Banyai \textit{et al.} [\onlinecite{banyai_prl98}]  which could be confirmed experimentally for optically excited semiconductors by Leitenstorfer \textit{et al.}~[\onlinecite{leitenstorfer_nature_01}].

This example also shows that, for times scales shorter than the correlation time, $t\lesssim \tau_{cor}$, the Balescu-Lenard equation is not applicable. Such time scales are easily accessible experimentally, e.g. in semiconductors. Moreover, this equation will also fail for strongly correlated systems because there the separation $t_{rel} \gg \tau_{cor}$ may be violated as well. As a consequence, the single-particle distribution and the pair correlation function may evolve on similar time scales. Mathematically, this means that Eq.~\eqref{eq:EOM:single-particle_fluctuations_PA(CP)_dissipative} will break down because it implies that 
\begin{align}
    \delta N_a(x,t)-\delta N^\mathrm{S}_a(x,t) \sim e^{-\delta(t-t_0)}\,,
\end{align}
where $t_0$ is the initial time. This expression vanishes if the initial time is shifted to the remote past, $t_0 \to -\infty$, which corresponds to Bogolyubov's condition of weakening of initial correlations and introduces irreversibility into the equations [\onlinecite{bonitz_98teubner,bonitz_cpp18}].

For physical processes at short time scales, on the order of the correlation time, $t-t_0 \sim \tau_{cor}$, in contrast, the time $t_0$ has to remain finite. The dynamics of the system then is a coupled time evolution of single-particle and two-particle quantities which must be complemented by initial conditions for both at the time $t_0$. In particular, one has to provide initial correlations.  This regime is properly captured by the G1--G2 scheme [\onlinecite{schluenzen_prl_20}] and also by the quantum fluctuations approach of the present authors -- a systematic generalization of Klimontovich's method to quantum systems that will be discussed in Sec.~\ref{s:quantum_fluctuations}.

\subsection{Further results and extensions  of the classical fluctuations approach} \label{ss:classical_fluctuations:extensions}
Similar as with the derivation of the Balescu-Lenard equation, Sec.~\ref{sss:ble}, Klimontovich applied his approach to derive a large variety of other kinetic equations, including equations for chemically reacting systems, relativistic equations, and non-Markovian equations for nonideal gases and plasmas. Moreover, he derived coarse grained equations such as hydrodynamic and gasdynamic equations and their generalizations to nonideal systems. A comprehensive overview has been given in his text book [\onlinecite{klimontovich_stat-phys}].

\subsubsection{Selection of further applications}\label{sss:further-applications}
Klimontovich's method has been picked up by various communities and applied to diverse problems. One example are dilute astrophysical plasmas such as the interstellar medium. There particle collisions are rare and electromagnetic field fluctuations play an important role. Here, Klimontovich's approach allowed for the derivation of kinetic equations that take into account the noise that is spontaneously emitted, e.g. from magnetic field fluctuations as well as thermal noise. An overview can by found in Ref.~[\onlinecite{kolberg_phys-rep_18}]. Another application of the formalism led V.~Belyi to an extension of the Balescu-Lenard kinetic equation and the involved dielectric function, cf. Sec.~\ref{sss:ble}, to strongly nonuniform systems where spatial gradients are important [\onlinecite{belyi_prl_02}]. Another field where Klimontovich's method turned out to be very effective is dusty plasmas. These are low-temperature and low-density plasmas that are dominated by neutrals and that contain micrometer size (``dust'') particles that charge up to thousands of elementary charges. Correspondingly, the dust particles may be strongly correlated, e.g. [\onlinecite{bonitz_rpp_10}] and exhibit liquid-like and even crystalline behavior, for a recent overview, see Ref.~[\onlinecite{dust-review_pop_23}]. The behavior of these systems is significantly influenced by fluctuations of the charge of the microparticles and by charging processes. Zagorodny et al. applied Klimontovich's approach to derive generalized kinetic equations for dusty plasmas [\onlinecite{zagorodny_prl_00,schram_pre_00}]. A similar approach is due to Tsytovich and co-workers who extended Klimontovich's phase space distribution by including the dust charge as an additional independent variable which led them to a generalization of the Balescu-Lenard equation [\onlinecite{tsytovich_pop_99}]. Finally, Tolias et al. computed the spectra of ion density and field fluctuations under the effect of dusty plasmas [\onlinecite{tolias_pre_12}], and Tolias recently also presented a detailed methodological analysis of Klimontovich's method in applications to dusty plasmas [\onlinecite{tolias_cpp_23}]. More information on the application of Klimontovich's method can be found in the overview by Bonitz and Zagorodny~[\onlinecite{zagorodny_cpp_24}].

\subsubsection{Fluctuation ensembles. Averaging procedure. Reversible vs. irreversible dynamics}\label{sss:averaging}
It is a bit surprising that, as far as we know, Klimontovich's fluctuations approach has, aside from formal derivations of approximations, practically not been applied to numerical solutions. In fact, the equations of motion of the fluctuations $\delta N$ or the correlation functions $\gamma_{ab}$ are well suited for numerical evaluation. 
We will show below, in Secs.~\ref{s:quantum_fluctuations} and \ref{s:stochastic_approach}, for the case of quantum systems, how such an approach can be straightforwardly developed that is capable to achieve results that are competitive with other methods.
A crucial basis for such a computational approach is physically based input about the underlying random process, i.e. about the ensemble of realizations of the fluctuations (of microstates).
Even though Klimontovich has not explicitly specified the ensemble of fluctuations, in most applications the derivations imply a thermodynamic equilibrium situation where field fluctuations, in general, are characterized by a Bose distribution. The only assumption that enters in the derivation of kinetic equation is Bogolyubov's condition of weakening of initial correlations. This is realized in the approach by introducing a small positive frequency correction $\delta$ in Eq.~\eqref{eq:EOM:single-particle_fluctuations_PA(CP)_dissipative}. This transforms the resulting kinetic equation (such as the Balescu-Lenard equation) into a dissipative one that is time irreversible. This means that pair correlations in the system have always achieved an equilibrium shape that may change slowly via the time dependent single-particle distributions, e.g. ~[\onlinecite{bonitz_98teubner}].

There are two ways to improve this result:
\begin{itemize}
    \item Avoid the complete neglect of initial correlations, but neglect only small scale correlations. Examples for long-lived or/and large scale fluctuations are chemical bound states, collective quantum states (e.g. quantum coherence), mesoscopic vortices, e.g. in the presence of turbulence and so on. Klimontovich has developed a heuristic concept of a subdivision of correlations into small-scale and large-scale correlations by introducing ``physically infinitesimal'' length and time scales $l_p \gg l_{cor}$ and $\tau_p \gg \tau_{cor}$. For the correlation length he uses a length scale $r_0$ which is the extension of an atom (or interaction range of a particle), whereas in a plasma the length scale is  the Debye screening length, $ r_D$. He uses the small parameter $l_{cor}/l_p = \epsilon :=\sqrt{nr_0^3} \ll 1$. This means that the mean interparticle distance is much smaller than the correlation length or that a sphere with radius $r_0$ contains a large number, $N_p \sim 1/\epsilon \gg 1$, of particles. Finally, all quantities are now averaged over a time scale $\tau_p$ and a length scale $l_p$, giving rise to smoothened functions $\tilde N$, $\tilde \gamma$,  $\tilde f_1$, $\tilde g_2$ and so on. As a result, one recovers the previous kinetic equation (e.g. the Balescu-Lenard equation) where $f_1$ is replaced by $\tilde f_1$ which do not contain information about the small scales, i.e. the atomic or molecular structure of the system anymore. In addition, however, the equation will contain an additional collision integral that is due to large scale correlations, $\tilde g_2$, e.g.~[\onlinecite{klimontovich_ufn_83}].
\item Another way to improve the kinetic equations that were discussed above is to drop the Bogolyubov condition of weakening of initial conditions entirely. While this condition appears to be natural for ``normal'' plasmas, where two-particle and single-particle time scales are well separated, $\tau_{cor}\ll t_{rel}$, one often recovers different situations. One case are strongly correlated systems. Another one are short time scales, $t\lesssim \tau_{cor}$, see our discussion in Sec.~\ref{sss:ble-short}. Such situations have recently attracted high interest in strongly correlated solids or cold atoms in optical lattices. These systems, at short times, behave as isolated systems without dissipative coupling to the environment.
In that case, it is not justified to introduce a frequency correction $\delta$ in Eq.~\eqref{eq:EOM:single-particle_fluctuations_PA(CP)_dissipative}. Instead, this equation has to be solved with an initial condition for the fluctuations and correlations at a finite initial time $t_0$. As a result, there appears a coupled system of equations for $f_1(t)$ and $g_2(t)$, e.g.~[\onlinecite{bonitz_98teubner, bonitz-etal.96pla}]. This system is time-reversible and contains collision effects. Such a system of equations was recently derived from nonequilibrium Green functions and was called G1--G2 scheme [\onlinecite{schluenzen_prl_20,joost_prb_20}]. An equivalent system can be derived using a quantum generalization of Klimontovich's fluctuations approach. This is explained in the next section.
\end{itemize}

%- In later works, on the other hand, Klimontovich concentrated on open systems,  selforganization processes and the connection between structure formation and dissipation. for details see Ref.~[\onlinecite{klimontovich_ufn_83}]

%\subsubsection{Classical fluctuations and Langevin approach} \label{sss:classical_fluctuations:Langevin}
%\subsubsection{Smoothing of classical fluctuations} \label{sss:classical_fluctuations:smoothing}
%\subsubsection{Creation and annihilation of particles \textcolor{red}{nötig?}} \label{sss:classical_fluctuations:finite_lifetime}
%\subsubsection{Product approach (?) \textcolor{red}{nötig?}} \label{sss:classical_fluctuations:prod(?)}

%\subsubsection{Discussion of the Averaging procedure}

\section{Quantum fluctuations approach} \label{s:quantum_fluctuations}
\subsection{Fluctuations of nonequilibrium Green functions} \label{ss:NEGF_fluctuations}
Generalizations of the classical theory of distribution functions for the description of quantum systems are given, for example, by the theories of reduced density matrices (RDM) and nonequilibrium Green functions (NEGF). Both can be described using the formalism of second quantization, which is characterized by the bosonic/fermionic creation ($\hat{c}^\dagger_i$) and annihilation ($\hat{c}_i$) operators on the so-called Fock space $\mathcal{F}$. More specifically, the Hilbert space $\mathcal{F}$ is induced by a single-particle Hilbert space $\mathcal{H}$, which has an orthonormal basis $(\psi_i)_i$. The creation operator $\hat{c}^\dagger_i$ then creates a particle in the orbital $\psi_i$ whereas the annihilation operator $\hat{c}_i$ annihilates a particle in said orbital. These operators obey the following (anti)commutation relations
\begin{equation}
    \Big[\hat{c}_i,\hat{c}^\dagger_j\Big]_\mp =\delta_{ij}\,,\quad \Big[\hat{c}_i,\hat{c}_j\Big]_\mp=\Big[\hat{c}^\dagger_i,\hat{c}^\dagger_j\Big]_\mp=0\,,\label{eq:properties_second_quantization}
\end{equation}
where the upper (lower) sign corresponds to bosons (fermions).\\
In this work, we consider a generic Hamiltonian of the following form
\begin{equation}
    \hat{H}\coloneqq \sum_{ij} h_{ij} \hat{c}^\dagger_i\hat{c}_j+\frac{1}{2}\sum_{ijkl}w_{ijkl}\hat{c}^\dagger_i\hat{c}^\dagger_j\hat{c}_l\hat{c}_k\,, \label{eq:definition:Hamiltonian}
\end{equation}
where $h_{ij}$ describes the single-particle contributions due to the kinetic energy and an external potential, and $w_{ijkl}$ describes two-particle interactions. Both contributions can, in general, be time-dependent, for example, to describe time-dependent excitations due to lasers or the creation of a correlated initial state from an uncorrelated state by adiabatically switching on the interaction. \\

A general quantum many-body system is determined by an ensemble of quantum states $(\Psi^{(k)})_k$ which form a complete orthonormal basis of the $N$-particle Hilbert space. Further, each of these states describing the microscopic configuration of the system is realized with a certain probability $p_k$, i.e., we have  $p_k \geq 0$ and $\sum_k p_k =1$. Analogously to the classical distribution function $P_N$, the quantum system can then be described using the $N$-particle density operator given by the sum of projections onto the vectors of the basis weighted by the probabilities, i.e., 
\begin{equation}
    \hat{\rho}_N\coloneqq \sum_k p_k |\Psi^{(k)}\rangle \langle \Psi^{(k)}|\,. \label{eq:definition:density_operator}
\end{equation}
As its classical analogue, the probability $P_N$, the $N$-particle density operator contains the full information about the quantum system and is invariant under the permutation of particles. Further, following from the Schrödinger equation of the $N$-particle states, the dynamics of the density operator is determined by the von Neumann equation, the quantum analogue of the Liouville equation, cf. Eq.~\eqref{eq:EOM:Liouville}, 
\begin{equation}
    \mathrm{i}\hbar\partial_t\hat{\rho}_N(t)=\left[\hat{H}(t), \hat{\rho}_N(t)\right]\,,
\end{equation}
with initial condition given by the initial states of the underlying quantum states $\Psi^{(k)}(t_0)$. It is important to highlight that it is assumed that the probabilities $p_k$ are time-independent. This is justified if, for example, the ``bath'', in which the system is embedded, is significantly larger than the subsystem.  \\
Moreover, given an observable $A$ described by an operator $\hat{A}$, its expectation value is given by
\begin{equation}
    \langle \hat{A}\rangle \coloneqq \mathrm{Tr}\left[ \hat{A}\hat{\rho}_N \right]\,.
\end{equation}
In general, however, a description of a quantum system by means of the $N$-particle density operator is out of reach and, similar to the classical case, it is possible to instead consider reduced quantities that contain the information of interest. These are given by the RDMs, defined in analogy to the $s$-particle distribution function $f_s$, cf. Eq.~\eqref{eq:definition:s-particle_distribution_function}, as
\begin{equation}
    F^{(s)}_{i_1\dots i_s j_1\dots j_s}(t)\coloneqq \frac{N!}{(N-s)!}\sum_{i_{s+1},\dots, i_N,j_{s+1},\dots, j_N} \rho_{i_1\dots i_si_{s+1}\dots i_Nj_1\dots j_sj_{s+1}\dots j_N}(t)\,,\label{eq:definition_s-particle_density_matrix}
\end{equation}
where $\rho$ on the r.h.s. denotes the density matrix associated with $\hat{\rho}$ and the basis $(\psi_i)_i$. Equivalently, the $s$-particle RDM can be expressed in the framework of second quantization in terms of the creation and annihilation operators, i.e., 
\begin{equation}
    F^{(s)}_{i_1\dots i_s j_1\dots j_s}(t)= \left\langle \hat{c}^\dagger_{j_1}(t)\dots\hat{c}^\dagger_{j_s}(t)\hat{c}_{i_s}(t)\dots \hat{c}_{i_1}(t)\right\rangle = \mathrm{Tr}\left[\hat{\rho} \hat{c}^\dagger_{j_1}(t)\dots\hat{c}^\dagger_{j_s}(t)\hat{c}_{i_s}(t)\dots \hat{c}_{i_1}(t)\right ] \,. \label{eq:s-particle_density_matrix_2}
\end{equation}
Analogously to the classical case, given an $s$-particle observable $A$, its expectation value is expressed via the $s$-particle RDM rather than the full $N$-particle density operator, i.e., 
\begin{equation}
    \langle \hat{A}\rangle =\frac{1}{s!} \mathrm{Tr}\left[A F^{(s)}\right]\coloneqq \frac{1}{s!}\sum_{i_1,\dots, i_s,j_1,\dots,j_s}A_{i_1\dots i_s j_1\dots j_s}F^{(s)}_{j_1\dots j_s i_1\dots i_s}\,.
\end{equation}
However, it can turn out to be useful to consider a generalization of the theory of RDM in the form of NEGF.
Here, the central quantity is given by the $s$-particle Green function on the Keldysh contour $\mathcal{C}$ defined as
\begin{equation}
    G^{(s)}_{i_1\dots i_s j_1\dots j_s}(z_1,\dots,z_s,z'_1,\dots z'_s)\coloneqq \left(\frac{1}{\mathrm{i}\hbar}\right)^s\left\langle \mathcal{T}_\mathcal{C}\left[ \hat{c}_{i_1}(z_1)\dots \hat{c}_{i_s}(z_s)\hat{c}^\dagger_{j_s}(z'_s)\dots \hat{c}^\dagger_{j_s}(z'_s)\right]\right\rangle \label{eq:definition:s-particle_NEGF}
\end{equation}
where $\mathcal{T}_\mathcal{C}$ denotes the time-ordering operator on the contour. Moreover, we write for the single-particle NEGF simply $G\coloneqq G^{(1)}$. The $s$-particle RDM can thus be considered a special equal-times limit of the $s$-particle NEGF that we denote as lesser component given by 
\begin{align}
    G^{(s),<}_{i_1\dots i_s j_1\dots j_s}(t_1,\dots,t_s,t_1',\dots,t_s')\coloneqq & \left(\pm\frac{1}{\mathrm{i}\hbar}\right)^s \left\langle \hat{c}^\dagger_{j_1}(t_1')\dots\hat{c}^\dagger_{j_s}(t_s')\hat{c}_{i_s}(t_s)\dots \hat{c}_{i_1}(t_1)\right\rangle \\
    G^{(s),<}_{i_1\dots i_s j_1\dots j_s}(t)\coloneqq & G^{(s),<}_{i_1\dots i_s j_1\dots j_s}(t,\dots,t)\,,\\
    G^{(s),<}_{i_1\dots i_s j_1\dots j_s}(t)= & \left(\pm\frac{1}{\mathrm{i}\hbar}\right)^s F_{i_1\dots i_s j_1\dots j_s}^{(s)}(t)\,.   
\end{align}
Analogously, it is possible to define greater components $G^{(s),>}$ by inverting the order of the operators and using a prefactor of $1/(\mathrm{i}\hbar)^s$ instead.\\
Hence, although reduced density matrices represent can be considered a more immediate generalization of the classical theory, the extension by means of the contour formalism proves to be particularly useful as it allows for the systematic derivation of approximations using, for example, Feynman's diagram approaches. However, many of these approximations can then in turn be translated to the framework of RDM, for example, using the G1--G2 scheme\cite{schluenzen_prl_20,joost_prb_20,joost_prb_22}. It is therefore useful for many theoretical considerations to set the framework of NEGF as a starting point, even when working within the context of RDM. \\

The quantum analogue of the classical fluctuations theory can then be constructed by considering the operator associated with the lesser component of the single-particle Green function and its fluctuations given by\footnote{We define fluctuations of observables $A$ of quantum systems as $\delta\hat{A}\coloneqq \hat{A}-\langle \hat{A}\rangle$.} 
\begin{gather}
    \hat{G}^<_{ij}(t)\coloneqq \pm\frac{1}{\mathrm{i}\hbar} \hat{c}^\dagger_j(t)\hat{c}_i(t)\,. \label{eq:definition:single-particle_Green_function_operator} \\
    \delta\hat{G}_{ij}(t)\coloneqq \hat{G}^<_{ij}(t)- G^<_{ij}(t)\,.\label{eq:definition:single-particle_fluctuations}
\end{gather}
Notice that due to the properties of the creation and annihilation operators, cf. \eqref{eq:properties_second_quantization}, fluctuations of the lesser and greater component coincide, thus justifying not differentiating between the two components in this situation. Further, we define $s$-particle fluctuations as correlation functions of single-particle fluctuations $\delta\hat{G}$
\begin{gather}
    L^{(s)}_{i_1\dots i_s j_1\dots j_s} (t_1,\dots, t_s)\coloneqq \left\langle \delta\hat{G}_{i_1j_1}(t_1)\dots \delta\hat{G}_{i_sj_s}(t_s)\right\rangle\,,\label{eq:definition:s-particle_fluctuations}\\
    L^{(s)}_{i_1\dots i_sj_1\dots j_s}(t)\coloneqq L^{(s)}_{i_1\dots i_sj_1\dots j_s}(t,\dots,t) \,,
\end{gather}
where we further denote two-particle fluctuations as $L\coloneqq L^{(2)}$. Two-particle fluctuations can be considered a special case of the exchange-correlation (XC) function within the general framework of NEGF defined as
\begin{equation}
    L_{ijkl}(z_1,z_2,z'_1,z'_2)\coloneqq G^{(2)}_{ijkl}(z_1,z_2,z'_1,z'_2)- G_{ik}(z_1,z'_1) G_{jl}(z_2,z'_2) \,,\label{eq:definition:XC_function}
\end{equation}
which describes all exchange and correlation contributions to the two-particle NEGF. Within the equal-times limit of NEGF corresponding to RDM, we introduce the correlated part of the two-particle NEGF analogously to the classical two-particle correlation function $g_s$, cf. Eq.~\eqref{eq:definition:classical_two-particle_correlations}, via
\begin{equation}
    G^{(2),<}_{ijkl}(t)\equiv G^<_{ik}(t) G^<_{jl}(t)\pm G^<_{il}(t) G^<_{jk}(t)+\mathcal{G}_{ijkl}(t)\,, \label{eq:definition:G2}
\end{equation}
where exchange contributions are present that are not included in the classical expression. The mean-field contributions, i.e., the first term on the r.h.s. of Eq.~\eqref{eq:definition:G2}, correspond to the two-particle RDM of a system that is described by the superposition of uncorrelated particles. Combining the exchange and correlation contributions on the r.h.s. of Eq.~\eqref{eq:definition:G2} illustrates a picture that resembles the classical case much more closely and leads to another special case of the XC function, cf. Eq.~\eqref{eq:definition:XC_function}.   Analogously, it is possible to define all correlated parts of all other $s$-particle NEGF on the time diagonal. For the three-particle correlated part, $\mathcal{G}^{(3)}$, we have the relation 
\begin{align}
    G^{(3),<}_{ijklmn}(t) =&\, G^<_{il}(t)G^<_{jm}(t)G^<_{kn}(t)+G^<_{im}G^<_{jn}(t)G^<_{kl}(t)+G^<_{in}(t)G^<_{jl}(t)G^<_{km}(t)\\
   & \pm G^<_{im}(t)G^<_{jl}(t) G^<_{kn}(t) \pm G^<_{in}(t) G^<_{jm}(t) G^<_{kl}(t) \pm G^<_{il}(t) G^<_{jn}(t) G^<_{km}(t)\\
    &+G^<_{il}(t) \mathcal{G}_{jkmn}(t)\pm G^<_{im}(t) \mathcal{G}_{jkln}(t)\pm G^<_{in}(t)\mathcal{G}_{jkml}(t)\\
    &+G^<_{jm}(t)\mathcal{G}_{ikln}(t)\pm G^<_{jl}(t)\mathcal{G}_{ikmn}(t)\pm G^<_{jn}(t)\mathcal{G}_{iklm}(t)\\
    &+G^<_{kn}(t)\mathcal{G}_{ijlm}(t)\pm G^<_{kl}(t)\mathcal{G}_{ijmn}(t)\pm G^<_{km}(t) \mathcal{G}_{ijln}(t)\\
    &+\mathcal{G}^{(3)}_{ijklmn}(t)\,. \label{eq:definition:G3}
\end{align}
However, if we now consider the second and third moment of the lesser component of the single-particle Green function operator, we find
\begin{gather}
    \left\langle\hat{G}^<_{ik}(t)\hat{G}^<_{jl}(t)\right\rangle = G^<_{ik}(t) G^<_{jl}(t) +L_{ijkl}(t) \,,\\
    \left\langle\hat{G}^<_{il}(t)\hat{G}^<_{jm}(t)\hat{G}^<_{kn}(t)\right\rangle = G^<_{il}(t) G^<_{jm}(t) G^<_{kn}(t)+ G^<_{il}(t) L_{jkmn}(t)+G^<_{jm}(t) L_{ikln}(t)+G^<_{kn}(t)L_{ijlm}(t)+L^{(3)}_{ijklmn}(t)
\end{gather}
which has the same structure as in the classical case, cf Eqs.~\eqref{eq:classical_second_moment} and \eqref{eq:classical_third_moment}, and does not explicitly include any exchange contributions as in Eqs.~\eqref{eq:definition:G2} and \eqref{eq:definition:G3}. In this sense, quantum fluctuations show a similar behavior as in the classical case. 
\subsection{Properties of quantum fluctuations} \label{ss:quantum_properties}
A key difference between classical and quantum fluctuations is that the latter do not obey certain exchange symmetries, i.e., we have $L_{ijkl}(t)\neq L_{jilk}(t)$, whereas in the classical case it holds $\gamma(x,x',t)=\gamma(x',x,t)$. Instead we find in the quantum case that 
\begin{equation}
    L_{ijkl}(t)-L_{jilk}(t)=\pm\frac{1}{\mathrm{i}\hbar}\left[\delta_{il}G^<_{jk}(t)-\delta_{jk}G^<_{il}\right]\,. \label{eq:exchange_property_L}
\end{equation}
Similarly, other exchange properties that the NEGF naturally possess, i.e., $\mathcal{G}_{ijkl}(t)=\pm\mathcal{G}_{jikl}(t)$, are not satisfied by fluctuations.\\
Using the properties of the creation and annihilation operators, cf.~Eq. \eqref{eq:properties_second_quantization}, the following relation holds 
\begin{equation}
     \left\langle\hat{G}^<_{ik}(t)\hat{G}^<_{jl}(t)\right\rangle = G^{(2),<}_{ijkl}(t)\pm \frac{1}{\mathrm{i}\hbar}\delta_{il}G^<_{jk}(t)\,, \label{eq:quantum_second_moment-two-particle_NEGF}
\end{equation}
similar to the classical case, cf. Eq.~\eqref{eq:classical_second_moment_two-particle_distribution_function}. However, for the relation between quantum correlations and fluctuations we find 
\begin{equation}
    L_{ijkl}(t)= \pm G^>_{il}(t)G^<_{jk}(t)+\mathcal{G}_{ijkl}(t)\,, \label{eq:relation_L-G2}
\end{equation}
which differs from the classical expression, cf. Eq.~\eqref{eq:classical_relation_gamma_g2}. It follows that quantum source fluctuations are defined as 
\begin{equation}
    L^0_{ijkl}(t)\coloneqq \pm G^>_{il}(t) G^<_{jk}(t)\,, \label{eq:definition_quantum_source_fluctuations}
\end{equation}
and describe particle-hole pair fluctuations. In the classical limit, $G^> \to 1$ and we recover the classical source fluctuations, $\gamma^S$.

For three-particle fluctuations and connected to the correlated part $\mathcal{G}^{(3)}$ by the following relation
\begin{align}
    L^{(3)}_{ijklmn}(t)=&\, \pm G^<_{jl}(t) \mathcal{G}_{ikmn}(t) \pm G^<_{km}(t)\mathcal{G}_{ijln}(t)+G^<_{kl}(t)\mathcal{G}_{ijmn}(t)\\
    &\pm G^>_{im}(t)\mathcal{G}_{jkln}(t)\pm G^>_{jn}(t) \mathcal{G}_{iklm}(t)+G^>_{in}(t)\mathcal{G}_{jklm}\\
    &+G^>_{im}(t)G^>_{jn}(t)G^<_{kl}(t)+G^>_{in}(t)G^<_{jl}(t) G^<_{km}(t)+\mathcal{G}^{(3)}_{ijklmn}(t)\,. \label{eq:relation_L3-G3}
\end{align}
Lastly, it holds for quantum fluctuations that their partial traces vanish like in the classical case, cf. Eq.~\eqref{eq:classical_trace}, 
\begin{equation}
    \sum_{i_k} L^{(s)}_{i_1\dots i_k \dots i_s j_1 \dots i_k\dots j_s} (t)=0
\end{equation}
for all $k=1,\dots, s$. Additionally, it is possible to consider other tensor contraction of $s$-particle fluctuations, which do not vanish. For example, for two-particle fluctuations it follows that 
\begin{equation}
    \sum_k L_{ikkj}(t) =\frac{1}{\mathrm{i}\hbar} N G^>_{ij}(t)- \left(G^>G^<\right)_{ij}(t)
\end{equation}
Furthermore, if the single-particle Hilbert space $\mathcal{H}$ is finite dimensional, i.e., $K\coloneqq \mathrm{dim}(\mathcal{H})<\infty$, we have
\begin{equation}
    \sum_k L_{kijk}(t)=\frac{1}{\mathrm{i}\hbar} (N\pm K)G^<_{ij}(t)- \left( G^>G^<\right)_{ij}(t)\,.
\end{equation}
These latter two contractions do not have an immediate classical analogue and constitute important conditions that approximations have to satisfy in order to enhance stability of time propagation in numerical solutions. 

\subsection{Dynamics of quantum many-body systems in terms of fluctuations} \label{ss:quantum_dynamics}
The EOM for $\hat{G}^<(t)$ directly follows from the Heisenberg equation for the creation and annihilation operators and is given by\footnote{We define the commutator of two $s$-particle quantities $A,B$ as $[A,B]_{i_1\dots i_sj_1\dots j_s}\coloneqq \sum_{k_1,\dots,k_s}[ A_{i_1\dots i_sk_1\dots k_s} B_{k_1\dots k_sj_1\dots j_s}-B_{i_1\dots i_sk_1\dots k_s}A_{k_1\dots k_sj_1\dots j_s}]$.}
\begin{equation}
    \mathrm{i}\hbar \partial_t \hat{G}^<_{ij}(t)= \left [\hat{h}^\mathrm{H},\hat{G}^<\right]_{ij}(t)\,, \label{eq:EOM:single-particle_NEGF_operator}
\end{equation}
where we introduced the Hartree Hamiltonian and  selfenergy (operator) 
\begin{gather}
    \hat{h}^\mathrm{H}_{ij}(t)\coloneqq h_{ij}+\hat{\Sigma}^\mathrm{H}_{ij}(t)\,.\label{eq:definition:Hartree_Hamiltonian}\\
    \hat{\Sigma}^\mathrm{H}_{ij}(t) \coloneqq \pm\mathrm{i}\hbar \sum_{kl} w_{ikjl} \hat{G}^<_{lk}(t)
\end{gather}
Analogously to the classical case, Eq.~\ref{eq:EOM:single-particle_distribution_function}, the EOM for the lesser component of the single-particle NEGF in the equal-times limit is given by 
\begin{equation}
    \mathrm{i}\hbar \partial_t G^<_{ij}(t) = \left[ h^\mathrm{H}, G^< \right]_{ij}(t)+\left\langle \left[\delta \hat{\Sigma}^\mathrm{H}, \delta \hat{G}\right]_{ij}(t)\right\rangle\,, \label{eq:EOM:G1}
\end{equation}
where the last term on the r.h.s. defines the (quantum) collision term, explicitly given by
\begin{gather}
        I_{ij}(t)\coloneqq \pm\mathrm{i}\hbar \sum_{klp}w_{iklp}L_{plkj}(t)\,, \label{eq:definition:quantum_collision_term}\\
    \left[I+I^\dagger\right]_{ij}(t)\equiv \left\langle\left[ \delta\hat{\Sigma}^\mathrm{H},\delta\hat{G} \right]_{ij}(t)\right\rangle \,.
\end{gather}
The EOM for single-particle fluctuations then immediately follows
\begin{equation}
    \mathrm{i}\hbar \partial_t \delta\hat{G}_{ij}(t) = \left[h^\mathrm{H}, \delta\hat{G}\right]_{ij}(t)+\left[\delta\hat{\Sigma}^\mathrm{H},G^<\right]_{ij}(t)+\delta\left\{\left[\delta\hat{\Sigma}^\mathrm{H},\delta\hat{G}\right]_{ij}\right\}(t)\,.\label{eq:EOM:quantum_single-particle_fluctuations}
\end{equation}
The EOM for $\delta\hat{G}$ can now be used to derive the EOMs for all $s$-particle fluctuations. For two-particle fluctuations, it follows 
\begin{equation}
    \mathrm{i}\hbar\partial_t L_{ijkl}(t) =\left[ h^{(2),\mathrm{H}},L \right]_{ijkl}(t)+\pi_{ijkl}(t)+ C_{ijkl}(t)\,,\label{eq:EOM:L}
\end{equation}
where we introduced the two-particle Hartree Hamiltonian defined as
\begin{equation}
    h^{(2),\mathrm{H}}_{ijkl}(t)\coloneqq h^\mathrm{H}_{ik}(t)\delta_{jl}+h^\mathrm{H}_{jl}(t)\delta_{ik} \label{eq:definition:two-particle_Hartree_Hamiltonian}
\end{equation}
and the (quantum) polarization contribution $\pi$ given by
\begin{align}
    \pi_{ijkl}(t)\coloneqq& \pm\mathrm{i}\hbar \sum_{pqr}\left\{ L_{rjpl}(t)\left[ w_{ipqr}G^<_{qk}(t)-w_{pqkr}G^<_{iq}(t) \right]-L_{iqkp}(t)\left[ w_{pjqr}G^<_{rl}(t)-w_{prql}G^<_{jr}(t) \right]\right\}\\
   =&-\hbar^2\sum_{pqrs}w_{pqrs}\left[ L_{iskq}(t)L_{jrlp}(t)-L_{rjpl}(t)L_{siqk}(t) \right] \,. \label{eq:definition:quantum_polarization_term}
\end{align}
The last term on the r.h.s. of Eq.~\eqref{eq:EOM:L} describes the coupling to three-particle fluctuations and is of the form
\begin{equation}
    C_{ijkl}(t)\coloneqq \pm\mathrm{i}\hbar\sum_{pqr}\left[ w_{ipqr}L^{(3)}_{rqjpkl}(t)+w_{pjqr}L^{(3)}_{iqrkpl}(t)-w_{pqkr}L^{(3)}_{irjpql}(t)-w_{pqrl}L^{(3)}_{ijrkqp}(t)  \right].\label{eq:definition:C3}
\end{equation}
\subsection{Approximations for quantum fluctuations} \label{ss:quantum_approximations}
All approximations of reduced density matrix theory or of the equal-times limit of NEGF can be directly reformulated in terms of fluctuations as well. In addition, it is possible to construct new approximations that are derived from the classical theory of fluctuations, thereby extending the arsenal of available many-body approximations. Of particular importance are those approximations that can be expressed in terms of single-particle fluctuations, $\delta \hat G$, because for them one can apply the highly efficient stochastic mean-field theory (SMF) 
%that allows for an approximate solution of EOM for single-particle fluctuations 
without the need to solve the equations of the  two-particle fluctuations, details will be discussed in Sec.~\ref{ss:SMF}.

\subsubsection{Approximations of moments} \label{sss:quantum_moments_approximation}
Analogous to the classical case, the simplest approximations one can consider are given by the approximations of moments. For the (quantum) approximation of first moments, it is again assumed all contributions due to fluctuations are negligible, i.e., the EOM for the single-particle Green function is given by 
\begin{equation}
    \mathrm{i}\hbar\partial_t G^{<}_{ij}(t)= \left[ h^\mathrm{H},G^< \right]_{ij}(t)\,,\label{eq:EOM:G1_1M}
\end{equation}
thus recovering the standard (quantum) Vlasov or Hartree equation. As this equation does not include any exchange effects, it is only applicable if those effects are negligible, for example, for systems that are weakly degenerate. Moreover, it has to be noted that fluctuations are always present, even for ideal systems as fluctuations are then solely determined by source fluctuations. In turn, the Hartree--Fock approximation can be recovered by said assumption, i.e., $L\approx L^0$, thus leading to 
\begin{equation}
    \mathrm{i}\hbar\partial_t G^{<}_{ij}(t)=\left[h^\mathrm{HF},G^{<}\right]_{ij}(t)\,,\label{eq:EOM:G1_HF}
\end{equation}
where we introduced the single-particle Hartree--Fock Hamiltonian given by
\begin{gather}
    h_{ij}^\mathrm{HF}(t)\coloneqq h_{ij}+\Sigma^\mathrm{HF}_{ij}(t)\,,\label{eq:definition:HF_Hamiltonian}\\
    \Sigma^\mathrm{HF}_{ij}(t)\coloneqq \pm\mathrm{i}\hbar\sum_{kl}w_{ikjl}^\pm G^<_{lk}(t)\,, \label{eq:definition:HF-selfenergy}
\end{gather}
where $w^\pm_{ijkl}\coloneqq w_{ijkl}\pm w_{ijlk}$ and $\Sigma^\mathrm{HF}$ denotes the Hartree--Fock selfenergy. \\
The (quantum) approximation of second moments is given by neglecting all contributions to the EOM for $L$ due to three-particle fluctuations, i.e., we have
\begin{equation}
    \mathrm{i}\hbar\partial_t L_{ijkl}(t)=\left[h^{(2),\mathrm{H}},L\right]_{ijkl}(t)+\pi_{ijkl}(t)\,.\label{eq:EOM:L_2M}
\end{equation}
Similarly to approximation of first moments, important effects are not properly taken into account as three-particle fluctuations are generally not negligible as can be seen when considering the relation between $L^{(3)}$ and $\mathcal{G}^{(3)}$, cf. Eq.~\eqref{eq:relation_L3-G3}, by setting all correlation contributions to zero. Then, contributions due to the appearing single-particle Green functions lead to non-vanishing three-particle fluctuations. Nonetheless, quantum effects are not entirely neglected as Pauli blocking is partially accounted for within the polarization contribution $\pi$ that includes second-order Born scattering terms.\\
At the level of single-particle fluctuations the approximation of second moments follows by neglecting all terms in the EOM, cf. Eq.~\eqref{eq:EOM:quantum_single-particle_fluctuations} that are quadratic in $\delta\hat{G}$, i.e., we have
\begin{equation}
    \mathrm{i}\hbar\partial_t \delta\hat{G}_{ij}(t)=\left[h^\mathrm{H},\delta\hat{G}\right]_{ij}(t)+\left[\delta\hat{\Sigma}^\mathrm{H},G^<\right]_{ij}(t)\,. \label{eq:EOM:quantum_single-particle_fluctuations_2M}
\end{equation}
Thus, it can be easily seen that only the first two approximations of moments can be expressed at the level of single-particle fluctuations.

\subsubsection{Quantum polarization approximation} \label{sss:QPA}
The quantum polarization approximation (QPA) is the natural extension of the (classical) polarization approximation, cf. Sec.~\ref{sss:classical_fluctuations:polarization_approximation}. Analogously to the classical case,  here not all contributions due to thee-particle fluctuations are neglected. Instead, weak coupling is assumed. In this case, two-particle correlations are significantly smaller than two-particle fluctuations and three-particle correlations are negligible, i.e., we have
\begin{gather}
    |\mathcal{G}_{ijkl}(t)|\ll |L_{ijkl}(t)|,\\
    |\mathcal{G}_{ijkl}(t)|\ll |L^0_{ijkl}(t)|,\\
    |\mathcal{G}^{(3)}_{ijklmn}(t)|\approx 0.
\end{gather}
Applying this approximation to the term describing the coupling to three-particle fluctuations, cf. Eq.~\eqref{eq:definition:C3}, by using the relation between three-particle fluctuations and correlations, leads to the following EOM 
\begin{equation}
    \mathrm{i}\hbar\partial_t L_{ijkl}(t)= \left[ h^{(2),\mathrm{HF}},L \right]_{ijkl}(t)+\pi^\pm_{ijkl}(t)\,,  \label{eq:EOM:L_QPA}
\end{equation}
where $h^{(2),\mathrm{HF}}$ denotes the effective two-particle Hartree--Fock Hamiltonian defined as
\begin{equation}
    h^{(2),\mathrm{HF}}_{ijkl}(t)\coloneqq h^\mathrm{HF}_{ik}(t)\delta_{jl}+h^\mathrm{HF}_{jl}(t)\delta_{ik},
\end{equation}
and $\pi^\pm$ the (anti)symmetric (quantum) polarization contribution given by 
\begin{equation}
    \pi^\pm_{ijkl}(t)\coloneqq \pm\mathrm{i}\hbar \sum_{pqr}\left\{ L_{rjpl}(t)\left[ w^\pm_{ipqr}G^<_{qk}(t)-w^\pm_{pqkr}G^<_{iq}(t) \right]-L_{iqkp}(t)\left[ w^\pm_{pjqr}G^<_{rl}(t)-w^\pm_{prql}G^<_{jr}(t) \right]\right\}\,.
\end{equation}
Essentially, the polarization approximation leads to the inclusion of further exchange contributions by using the (anti)symmetrized interaction $w^\pm$ instead of the pair interaction $w$. In contrast to the Hartree--Fock approximation at the one-particle level, however, this does not correspond to a description of all exchange effects at the level of two-particle fluctuations. For this, it would also be necessary to consider a contribution of the form $P_{ijkl}(t)\coloneqq \pi^\pm_{ijkl}(t)\pm\pi^\pm_{ijlk}(t)$ instead of the (anti)symmetric (quantum) polarization contribution. \\
Additionally, it is also necessary to consider the EOM for source fluctuations that follows from the EOM for the single-particle Green function, cf. Eq.~\eqref{eq:EOM:G1} and is given by
\begin{gather}
    \mathrm{i}\hbar\partial_t L^0_{ijkl}(t)= \left[ h^{(2),\mathrm{H}},L^0 \right]_{ijkl}(t)+R_{ijkl}(t)\,, \label{eq:EOM:L0}\\
    R_{ijkl}(t)\coloneqq \pm\left\{G^>_{il}(t)\left[I+I^\dagger\right]_{jk}(t)+\left[I+I^\dagger\right]_{il}(t)G^<_{jk}(t) \right\}\,.
\end{gather}
Applying the QPA to the EOM for source fluctuations leads to 
\begin{equation}
   \mathrm{i}\hbar\partial_t L^{0,\mathrm{P}}_{ijkl}(t)= \left[ h^{(2),\mathrm{HF}},L^{0,\mathrm{P}} \right]_{ijkl}(t)\,. \label{eq:EOM:L0_QPA}
\end{equation}
Using the EOMs for $L$ and $L^{0,\mathrm{P}}$, cf. Eqs.~\eqref{eq:EOM:L_QPA} and \eqref{eq:EOM:L0_QPA}, it can then be shown that the QPA is equivalent to the $GW$ approximation within the G1--G2 scheme with additional exchange contributions included in the weak coupling limit\footnote{These additional exchange contributions corresponds to those arising from the replacement $w\rightarrow w^\pm$ in the polarization term $\pi$. Thus, equivalence to the $GW$ approximation can be obtained by considering $\pi$ instead of $\pi^\pm$ in Eq.~\eqref{eq:EOM:L_QPA}.}, cf. Ref.~ [\onlinecite{schroedter_cmp_22}].\\
At the level of single-particle fluctuations the QPA follows from the approximation
\begin{equation}
    \delta\hat{G}_{ik}(t)\delta\hat{G}_{jl}(t)-\left\langle \delta\hat{G}_{ik}(t)\delta\hat{G}_{jl}(t)\right\rangle \approx \pm\left[ G^>_{il}(t)\delta\hat{G}_{jk}(t)+\delta\hat{G}_{il}(t)G^<_{jk}(t) \right].
\end{equation}
While the l.h.s. of the equation describes fluctuations of two-particle fluctuations, the r.h.s. can be considered the equivalent of fluctuations of source fluctuations. Hence, at the level of single-particle fluctuations, the QPA can be considered the analogue of the Hartree--Fock approximation for $G^<$. Consequently, the EOM for single-particle fluctuations is given by
\begin{equation}
    \mathrm{i}\hbar\partial_t \delta\hat{G}_{ij}(t)=\left[h^\mathrm{HF},\delta\hat{G}\right]_{ij}(t)+\left[\delta\hat{\Sigma}^\mathrm{HF},G^<\right]_{ij}(t)\,. \label{eq:EOM:quantum_single-particle_fluctuations_QPA}
\end{equation}

\subsubsection{Exchange-correlation function depending on two times} \label{sss:l-2times}
In the following, we will also be interested in dynamic (frequency-dependent) observables. This is achieved as in the classical case by considering exchange-correlation functions depending on two times. Their equations of motion follow from Eq.~\eqref{eq:EOM:quantum_single-particle_fluctuations_QPA}, and Eq.~\eqref{eq:EOM:L_QPA},
\begin{align}
    \mathrm{i}\hbar\partial_{t_m} L_{ijkl}(t_1,t_2) &= \left[h^\mathrm{HF},L\right]^{(m)}_{ijkl}(t_1,t_2)+\pi^{(m),\pm}_{ijkl}(t_1,t_2)\,,
\end{align}
for $m=1,2$. Due to the symmetry $\delta\hat{G}_{ij}(t)=-[\delta\hat{G}_{ji}(t)]^\dagger$ and thus $L_{ijkl}(t_1,t_2)=\left[L_{lkji}(t_2,t_1)\right]^*$, it is sufficient to only consider the EOM for $m=1$ as the EOM for $m=2$ follows analogously as the contributions obey the symmetries:
\begin{gather}
    \left[h^\mathrm{HF},L\right]_{ijkl}^{(1)}(t_1,t_2)=-\left\{\left[h^\mathrm{HF},L\right]_{lkji}^{(2)}(t_1,t_1)\right\}^*\,,\\
    \pi^{(1),\pm}_{ijkl}(t_1,t_2)=-\left\{\pi^{(2),\pm}_{lkji}(t_2,t_1)\right\}^*\,.
\end{gather}
The two-time Hartree--Fock and polarization contributions are given by
\begin{gather}
    \left[h^\mathrm{HF},L\right]_{ijkl}^{(1)}(t_1,t_2)\coloneqq \sum_p\left[ h^\mathrm{HF}_{ip}(t_1)L_{pjkl}(t_1,t_2)- h_{pk}^\mathrm{HF}(t_1)L_{ijpl}(t_1,t_2)\right]\,,\\
    \pi^{(1),\pm}_{ijkl}(t_1,t_2)\coloneqq \pm\mathrm{i}\hbar\sum_{pqr} L_{rjpl}(t_1,t_2)\left[ w^\pm_{ipqr}G^<_{qk}(t_1)-w^\pm_{qpkr}G^<_{iq}(t_1) \right]\,.
\end{gather}
Most importantly, the relation of the QPA and the $GW$ approximation extends from the single-time case to the corresponding two-time version of both approximations. \\
A two-time formulation turns out to be particularly useful as a variety of observables, that are experimentally accessible, depend on two-time fluctuations such as the response functions and their corresponding structure factors, e.g., the retarded component of the density response function is given by 
\begin{equation}
    \chi^\mathrm{R}_{ij}(t_1,t_2) \coloneqq \mathrm{i}\hbar\Theta(t_1-t_2)\left\langle\left[\delta\hat{G}_{ii}(t_1),\delta\hat{G}_{jj}(t_2)\right]\right\rangle \,. \label{eq:definition:Chi}
\end{equation}

\section{Stochastic approach to quantum fluctuations} \label{s:stochastic_approach}
\subsection{Stochastic mean-field theory} \label{ss:SMF}
Within the quantum theory of fluctuations, single-particle fluctuations are operators on a Fock space and, thus, their EOMs, cf. Eq.~\eqref{eq:EOM:quantum_single-particle_fluctuations}, and even approximations, e.g., Eq.~\eqref{eq:EOM:quantum_single-particle_fluctuations_QPA}, are rarely solvable. A possibility to circumvent this issue is given in the form of stochastic mean-field (SMF) theory as developed by Ayik\cite{ayik_plb_08} and later extended by Lacroix and many others, see, e.g., \onlinecite{lacroix_prb14,lacroix_epj_14,Yilmaz2014,Czuba2020}. Here, similar to other semiclassical approaches like the truncated Wigner approximation (TWA), operators are replaced by random variables, i.e., $\delta\hat{G}_{ij}(t)\rightarrow \Delta G^\lambda_{ij}(t)$\footnote{Only replacing the single-particle fluctuations operator with an ensemble of realizations turns out to be sufficient to describe the fluctuations of any relevant operator since, in second quantization, they can be expressed in terms of single-particle Green functions and fluctuations.}, and the quantum-mechanical expectation value by its stochastic semiclassical counterpart, i.e., $\langle\cdot\rangle\rightarrow \overline{(\cdot)}$. More precisely, this means that is assumed that there exists a probability distribution $\mathcal{P}_N$, similar to the classical case with its distribution $P_N$, which describes the statistical properties of the fluctuations in the quantum system. The microstate $\Delta G^\lambda$ is then associated with a random realization of the system. Furthermore, since our considerations are done within the Heisenberg picture, the distribution $\mathcal{P}_N$ itself has no time dependence and instead the dynamical properties of the system are determined by the propagation of the realizations of the ensemble of fluctuations. \\
%Given a single-particle observable $A$, 
%\begin{equation}
%    \Delta A^\lambda(t) \coloneqq \pm\mathrm{i}\hbar\mathrm{Tr}\left[A\Delta G^\lambda(t)\right]= \pm\mathrm{i}\hbar \sum_{ij} A_{ij}\Delta G^\lambda_{ji}(t)
%\end{equation}
%\begin{equation}
%    \overline{(\Delta A^\lambda(t))^2}=-\hbar^2 \sum_{ijkl} A_{ij}A_{kl} \overline{\Delta G^\lambda_{ji}(t)\Delta G^\lambda_{lk}(t)}
%\end{equation}
In practical application, it is generally necessary to approximate the stochastic expectation value by the arithmetic mean instead of using the probability distribution $\mathcal{P}_N$. Then, the superscript '$\lambda$' denotes a random realization of the ensemble. Within SMF theory, this ensemble is generated such that it corresponds to the initial state of the quantum system. Each realization is then propagated in time using the EOM for single-particle fluctuations, e.g., Eq.~\eqref{eq:EOM:quantum_single-particle_fluctuations}. Here, however, it has to be underlined that, replacing \textit{noncommuting} operators with random variables, requires special care, as random classical variables commute. An attempted solution to this problem is given by replacing only symmetrized products of operators with random variables, i.e., 
\begin{equation}
    \frac{1}{2}\left[ \delta\hat{G}_{ij}(t)\delta\hat{G}_{kl}(t)+\delta\hat{G}_{kl}(t)\delta\hat{G}_{ij}(t) \right]\rightarrow \Delta G^\lambda_{ij}(t)\Delta G^\lambda_{kl}(t)\,.
\end{equation}
Consequently, the ensemble of random realizations is constructed such that all \textit{symmetrized} moments of the ideal quantum initial state are correctly reproduced, i.e., we have for the first two moments
\begin{align}
    \overline{\Delta G^\lambda_{ij}(t_0)} &= 0\,,\label{eq:first_moment_SMF}\\
    \overline{\Delta G^\lambda_{ij}(t_0)\Delta G^\lambda_{kl}(t_0)}&= -\frac{1}{2\hbar^2} \delta_{il}\delta_{jk}\left[ n_j(1\pm n_i)+n_i(1\pm n_j) \right]\,, \label{eq:second_moment_SMF}
\end{align}
where $n_i\coloneqq F^{(1)}_{ii}(t_0)$. Only considering an ideal initial state does not constitute a restriction of this approach as a correlated initial state can be generated from an ideal state using the adiabatic switching method\cite{schluenzen_cpp16}. Knowledge of all moments should, in theory, uniquely determine the probability distribution $\mathcal{P}_N$ describing the initial configuration of the system. However, this is not the case as it is, in general, impossible for such a probability distribution to exist\cite{Lacroix2019}. This will be discussed in more detail in Sec.~\ref{ss:sampling}.\\
Within the SMF framework, the dynamics of the quantum system is described in terms of the single-particle Green function and the ensemble of microstates given by the fluctuations. Then, the following set of equations describes the evolution of the system:
\begin{gather}
    \mathrm{i}\hbar\partial_t G^<_{ij}(t)= \left[h^\mathrm{H},G^<\right]_{ij}(t)+\left[S+S^\dagger\right]_{ij}(t)+\overline{\left[ \Delta\Sigma^{\mathrm{H},\lambda},\Delta G^{\lambda} \right]}_{ij}(t)\,, \label{eq:EOM:G1-SMF}\\
    \mathrm{i}\hbar\partial_t\Delta G^\lambda_{ij}(t)=\left[h^\mathrm{H},\Delta G^\lambda\right]_{ij}(t)+\left[ \Delta\Sigma^{\mathrm{H},\lambda}, G^< \right]_{ij}(t)+\left[ \Delta\Sigma^{\mathrm{H},\lambda},\Delta G^{\lambda} \right]_{ij}(t)-\overline{\left[ \Delta\Sigma^{\mathrm{H},\lambda},\Delta G^{\lambda} \right]}_{ij}(t)\,, \label{eq:EOM:single-particle_fluctuations_SMF}
\end{gather}
where $\Delta\Sigma^{\mathrm{H},\lambda}$ denotes the realization of the fluctuations Hartree selfenergy $\delta\hat{\Sigma}^\mathrm{H}$ and $S$ describes a contribution that arises due to the symmetrization of the collision term, cf. Eq.~\eqref{eq:definition:quantum_collision_term}, and is given by
\begin{equation}
    S_{ij}(t)\coloneqq \frac{1}{2}\sum_{kl}w_{kljk}G^<_{il}(t)\,. \label{eq:definition:S-term}
\end{equation}
The dynamics of the single-particle Green function or the single-particle density is determined on the one hand by the interaction with the mean field of the particles and the ensemble averaged interaction of the mean-field of the fluctuations with themselves. The dynamics of each realization of a microstate of the ensemble is then determined by a combination of the interaction of the state with the mean field of the particles and the interaction of the mean field of the microstate with the particles. In addition, there are contributions describing the fluctuations of the interaction of the individual realization of the fluctuations' mean field and the fluctuations.\\
Usually, within SMF theory, instead of considering the set of coupled equations given by Eqs.~\eqref{eq:EOM:G1-SMF} and \eqref{eq:EOM:single-particle_fluctuations_SMF}, the SMF version of the EOM for the single-particle Green function operator $\hat{G}^<$, cf. Eq.~\eqref{eq:EOM:single-particle_NEGF_operator}, is used, as the arising equations are simple mean-field equations that can be propagated independently. Since equations of this kind are reversible, it follows that, applying the SMF approach in this form preserves the reversibility of the equations of motion which is similar to the reversibility of the G1--G2 scheme. Furthermore, it has to be mentioned that, although Eqs.~\eqref{eq:EOM:G1-SMF} and \eqref{eq:EOM:single-particle_fluctuations_SMF} provide a closed set of equations, their solution corresponds to the solution of an entire hierarchy of equations,  similar to the fluctuations hierarchy~[\onlinecite{Lacroix2016}]. This is, however, associated with several obstacles that will be discussed in more detail in the following section, cf. Sec.~\ref{ss:sampling}.

\subsection{Probability distributions and sampling}\label{ss:sampling}
The underlying assumption of SMF theory is the existence of a probability distribution $\mathcal{P}_N$ that can reproduce the initial quantum state. Hence, it should reproduce all moments of single-particle fluctuations. For aforementioned reasons, only symmetric moments can be considered. However, the questions remains whether such a distribution function exists. In principle, knowledge of all moments allows the corresponding probability distribution to be reconstructed. Therefore, one might assume that, given an initial state described by $\hat{\rho}(t_0)$, the associated moments would give rise to a unique probability distribution. However, this is, in general, not the case. Following the derivation given in Ref.~\onlinecite{Lacroix2019}, we consider a system of fermions at zero temperature, i.e., we have $n_i\in\{0,1\}$. For simplicity, we set $\Delta n^\lambda_{ij}\coloneqq \pm\mathrm{i}\hbar\Delta G^\lambda_{ij}(t_0)$. Then, the second moment, cf. Eq.~\eqref{eq:second_moment_SMF}, implies $\overline{|\Delta n_{ii}^\lambda|^2}=0$ and thus $\Delta n^\lambda_{ii}=0$. Further, it follows from Eq.~\eqref{eq:second_moment_SMF} all matrix entries are uncorrelated.\\
Let us now consider the third and fourth symmetric moment that are given by
\begin{align}
    \overline{\Delta n^\lambda_{ij}\Delta n^\lambda_{kl}\Delta n^\lambda_{mn}}&= \delta_{il}\delta_{jm}\delta_{kn} \Lambda^{(3)}_{jik}\,, \label{eq:third_moment_SMF}\\
    \overline{\Delta n^\lambda_{ij}\Delta n^\lambda_{kl}\Delta n^\lambda_{mn}\Delta n^\lambda_{pq}}&= \delta_{il}\delta_{kn}\delta_{qm}\delta_{jp} \Lambda^{(4,1)}_{jikq}+\delta_{il}\delta_{jk}\delta_{qm}\delta_{pn}3 \Lambda^{(4,2)}_{jipq} \label{eq:fourth_moment_SMF}\,,
\end{align}
where we introduced 
\begin{align}
    \Lambda^{(3)}_{ijk} \coloneqq &\frac{1}{3}\big[n_i(1- 3n_j)(1- n_jn_k)+n_k(1- 3n_i)(1- n_in_j)+n_j(1- 3 n_k)(1- n_in_k)  \big]\,,\\
    \Lambda^{(4,1)}_{ijkl} \coloneqq &\frac{1}{4} \big[ n_i(1- 4n_j)(1-3 n_k)(1-n_jn_kn_l)+n_k(1-4n_i)(1-3n_j)(1-n_in_jn_k) \\
    &\quad +n_k(1-4n_l)(1-3n_i)(1-n_in_jn_l)+n_j(1-4n_k)(1-3n_l)(1-n_in_kn_l)  \big]\,,\\
    \Lambda^{(4,2)}_{ijkl}\coloneqq &\frac{1}{4}\big[ n_in_k(1-n_jn_l)(1-n_j-n_l)+n_jn_k(1-n_in_l)(1-n_i-n_l)\\
    &\quad +n_in_l(1-n_jn_k)(1-n_j-n_k)+n_jn_l(1-n_in_k)(1-n_i-n_k)\big]\,.
\end{align}
Let $\mathscr{P}\coloneqq \{i\mid n_i=1\}$ denote the set of occupied states and $\mathscr{H}\coloneqq \{i \mid n_j=0\}$ the set of unoccupied states. Then, using the Kronecker deltas in Eq.~\eqref{eq:third_moment_SMF}, we have
\begin{align}
    \overline{\Delta n^\lambda_{ij}\Delta n^\lambda_{ki}\Delta n^\lambda_{jk}} = \begin{cases} \frac{1}{3}\,, & \mbox{if } (i,j,k)\in (\mathscr{P}\times \mathscr{H}^2)\cup (\mathscr{H}\times \mathscr{P}\times \mathscr{H})\cup (\mathscr{H}^2\times \mathscr{P}) \\
        -\frac{1}{3}\,, &\mbox{if } (i,j,k) \in (\mathscr{H}\times\mathscr{P}^2)\cup (\mathscr{P}\times \mathscr{H}\times\mathscr{P}) \cup (\mathscr{P}^2\times \mathscr{H})\\
        0\,, & \mbox{else}\,,
    \end{cases} \label{eq:third_moment_SMF_1}
\end{align}
i.e., the third moment is only nonzero if two of the involved states are either occupied and the other one empty or two of them are empty and the other one occupied. Further, we have for the fourth moment (again only considering a subset of indices due to the Kronecker deltas in Eq.~\eqref{eq:fourth_moment_SMF})
\begin{align}
    \overline{\Delta n^\lambda_{ij}\Delta n^\lambda_{ji}\Delta n^\lambda_{qp}\Delta n^\lambda_{pq}}&= \begin{cases}
        \frac{3}{4} \,, & \mbox{if } (i,j,p,q)\in ( \mathscr{P}\times \mathscr{H}^2 \times \mathscr{P}) \cup  \times (\mathscr{P} \times \mathscr{H}\times \mathscr{P}\cup \mathscr{H}) \cup (\mathscr{P}\leftrightarrow \mathscr{H}) %HPPH PHHP PHPH HPHP
        \\
        0\,, & \mbox{else}\,,
    \end{cases}\label{eq:fourth_moment_SMF_1}\\
    \overline{\Delta n^\lambda_{ij}\Delta n^\lambda_{ki}\Delta n^\lambda_{qk}\Delta n^\lambda_{jq}} &= \begin{cases}
        \frac{1}{4}\,, & \mbox{if } (i,j,k,q)\in (\mathscr{P}^3\times \mathscr{H})\cup (\mathscr{P}^2\times \mathscr{H}\times \mathscr{P})\cup (\mathscr{P}\times \mathscr{H}\times \mathscr{P}^2)\cup ( \mathscr{H}\times \mathscr{P}^3) \cup ( \mathscr{P}\leftrightarrow \mathscr{H}) \\
        -\frac{1}{2}\,, & \mbox{if } (i,j,k,q)\in (\mathscr{P}^2\times \mathscr{H}^2) \cup (\mathscr{P}\times\mathscr{H}\times\mathscr{P}\times\mathscr{H}) \cup (\mathscr{P}\leftrightarrow\mathscr{H})\,,\\
        -1\,, &\mbox{if } (i,j,k,q)\in (\mathscr{P}\times \mathscr{H}^2\times\mathscr{P})\cup (\mathscr{P}\leftrightarrow \mathscr{H})\,,\\
        0\,, &\mbox{else}\,.
    \end{cases} \label{eq:fourth_moment_SMF_2}
\end{align}
Equations~\eqref{eq:third_moment_SMF_1}, \eqref{eq:fourth_moment_SMF_1} and \eqref{eq:fourth_moment_SMF_2} imply that the entries are correlated. Moreover, these equations also admit nonzero contributions involving $\Delta n^\lambda_{ii}$, i.e., we have $\overline{\Delta n_{ij}^\lambda \Delta n^\lambda_{ji}\Delta n^\lambda_{jj}}=\Lambda^{(3)}_{jij}\neq 0$ and $\overline{\Delta n^\lambda_{ij}\Delta n^\lambda_{ii}\Delta n^\lambda_{ji}\Delta n^\lambda_{jj}}=\Lambda^{(4,1)}_{jiij}\neq 0$. Thus, in general, it is not possible for a probability distribution $\mathcal{P}_N$  to exist that satisfies the given constraints.\\
As stated before, the set of equations given by Eqs.~\eqref{eq:EOM:G1-SMF} and \eqref{eq:EOM:single-particle_fluctuations_SMF} corresponds to a hierarchy of equations that closely resembles the fluctuations hierarchy, cf. Sec.~\ref{ss:quantum_dynamics}. Although the full hierarchy is being solved within the framework of the SMF approach, the initial conditions are given by all moments. Hence, due to the nonexistence of a probability distribution that exactly reproduces all quantum moments, the propagation differs from the exact dynamics due to deviations in the initial conditions. Within SMF theory, it is therefore of interest to consider suitable probability distributions that minimize the deviations for higher moments. A different option, however, is given by considering further approximations within the SMF framework to mitigate the effect of higher moments on the propagation.
%\textcolor{red}{den nächsten Absatz in die folgende subsection integrieren. Dort kommen alle Infos zu den Ensembles und Verteilungen}
\\

\subsubsection{Stochastic sampling} \label{sss:stochastic_sampling}
The standard approach within SMF theory for the construction of the ensemble is given by stochastic sampling, i.e., random realizations of the initial state are generated according to a known probability distribution. Due to the appearing Kronecker deltas in Eq.~\eqref{eq:second_moment_SMF}, the second moment for $i=k\neq j=l$ vanishes, i.e., $\overline{(\Delta n_{ij}^\lambda)^2}=0$. Thus, in order to fulfill this condition, it is necessary to consider probability distributions of complex random variables. Further, as only the first two moments are considered, it is possible to sample a subset of matrix entries of $\Delta n$ independently as for any entry $\Delta n^\lambda_{ij}$ the entry $\Delta n^\lambda_{ji}$ is given by $\Delta n^\lambda_{ji}=[\Delta n_{ij}^\lambda]^*$ for each random sample. In order to ensure that the arithmetic mean corresponding to the first moment of fluctuations vanishes, for every sample $\Delta n_{ij}^\lambda$ a sample $-\Delta n_{ij}^\lambda$ is generated. Moreover, corresponding to the second moment of fluctuations, cf. Eq.~\eqref{eq:second_moment_SMF}, the variance of the each entry is given by 
\begin{equation}
    \sigma^2_{ij}\coloneqq \frac{1}{2}\left[ n_i(1\pm n_j)+n_j(1\pm n_i) \right] \geq0\,.\label{eq:definition:variance}
\end{equation}
 \\
The most common choice for a probability distribution is given by a complex Gaussian distribution of the following form:
\begin{equation}
    P_{ij}^\mathrm{N}(x,y)\coloneqq \frac{1}{\pi \sigma_{ij}^2}\exp\left(-\frac{x^2+y^2}{\sigma^2_{ij}}\right)\,,
\end{equation}
i.e., the real and imaginary part of $\Delta n_{ij}$ are sampled independently. Alternatively, it is possible to use a complex generalization of a two-point distribution, i.e., a four-point distribution, given by 
\begin{equation}
    P^{4\mathrm{P}}_{ij}(x,y)\coloneqq \frac{1}{4}\left\{ \delta(y)[\delta(x+\sigma_{ij})+\delta(x-\sigma_{ij})]+\delta(x)[\delta(y+\sigma_{ij})+\delta(y-\sigma_{ij})]\right\}\,.
\end{equation}
Both distributions reproduce the first and second moment, cf. Eq.~\eqref{eq:first_moment_SMF} and \eqref{eq:second_moment_SMF}, however, they fail to correctly reproduce higher moments. Further, higher moments of the complex two-point distribution deviate less from the exact moments compared to the complex Gaussian distribution. There have been extensive tests for the Lipkin--Meshkov--Glick model within the SMF approach\cite{Lacroix2019}. Here, it was observed what effects the choice of distribution has on the dynamics of the system and it has been shown that the complex two-point distribution lead to the best agreement with the exact solution. Additionally, quasiprobability distributions have been investigated within the context of the SMF approach and have been shown to further increase the accuracy of the results as it is then possible to further minimize the error with respect to the higher moments\cite{Yilmaz2014}. \\
Stochastic sampling of the initial state requires a sufficiently large number of samples to guarantee converged results. Numerical experiments, however, have shown that the number of required does not increase for larger systems and can thus be chosen constant\footnote{Generally, a number of about $10^4$ samples is required.}\cite{schroedter_cmp_22}. Consequently, this approach to sampling turns out to particularly efficient for very large systems. Numerical illustrations will be presented in Sec.~\ref{s:numerics}.

\subsubsection{Deterministic sampling} \label{sss:deterministic_sampling}
As the numerical scaling of simulations depends linearly on the number of samples, it is of great interest to develop other methods that require a small number of samples and, at the same time, guarantee converged results. The idea of deterministic sampling is to interpret the equations for the different moments as a system of nonlinear equations, i.e., 
\begin{align}
    \sum_{\lambda = 1}^M \Delta n_{ij}^\lambda &= 0\,,\label{eq:deterministic_sampling_1}\\
    \sum_{\lambda =1}^M \Delta n_{ij}^\lambda \Delta n^\lambda_{kl}&=M\sigma_{ij}^2 \delta_{il}\delta_{jk}\,,\label{eq:deterministic_sampling_2}
\end{align}
where $M$ is a parameter that has to be chosen such that a solution exists. It is therefore easily seen that this parameter generally depends on the number of basis states of the single-particle Hilbert space. \\
Here, we present an algorithm for the solution of this problem that was proposed in Ref.~\onlinecite{schroedter_cmp_22} for the special case of electrons at zero temperature, i.e, we have $n^\sigma_i\in\{0,1\}$ and $\sigma^{\sigma\sigma'}_{ij}=\frac{1}{2}(1- \delta_{n_i^\sigma n_j^{\sigma'}})$. Further, we assume a spin-symmetric initial state, i.e., $n_i^\uparrow = n_i^\downarrow$. Let $N_\mathrm{p}$ denote the number of particles with spin $\uparrow$. Without loss of generality, we have $n_i^\sigma =1$ for $i=1,\dots, N_\mathrm{p}$ and $n_i^\sigma=0$ for $i= N_\mathrm{p}+1,\dots ,N'_\mathrm{b}$, where $N'_\mathrm{b}$ denotes the size of the single-particle basis for one spin component. Further, we set $N_\mathrm{h}\coloneqq N'_\mathrm{b}-N_\mathrm{p}$, i.e., $N_\mathrm{h}$ denotes the number of holes for one spin component in the system. The solution of the system of nonlinear equations, cf. Eqs.~\eqref{eq:deterministic_sampling_1} and \eqref{eq:deterministic_sampling_2}, can thus be expressed by a set of matrices of the form
\begin{equation}
    \Delta n^{\lambda,\sigma} = \begin{pmatrix}
        0 & A^{\lambda, \sigma} \\ {A^{\lambda,\sigma}}^\dagger & 0 
    \end{pmatrix}\,,
\end{equation}
where $A^{\lambda, \sigma}\in\mathbb{C}^{N_\mathrm{h}\times N_\mathrm{p}}$.  If we now set $M= 8N_\mathrm{p}N_\mathrm{h}$, a solution can be constructed in the following way. Let $\varphi :\{1,\dots, N_\mathrm{h}\}\times\{1,\dots, N_\mathrm{p}\}\rightarrow\{1,\dots, N_\mathrm{p}N_\mathrm{h}\}$  be one-to-one. Then, by identifying $\{1,\dots, M\}\ni\lambda\leftrightarrow (\alpha,\beta)\in\{1,\dots, N_\mathrm{p}N_\mathrm{h}\}\times \{1,\dots, 8\}$, we have a solution of the form
\begin{align}
    A^{(\alpha,\beta),\sigma}_{ij}=\begin{cases}
        \mathrm{i}^\beta \sqrt{N_\mathrm{p}N_\mathrm{h}}\,, & \mbox{if either } (\alpha,\beta)\in \{ \varphi(i,j)\}\times\{ 1,\dots, 4\}, \sigma =\uparrow \mbox{ or } (\alpha,\beta)\in \{ \varphi(i,j)\}\times\{ 5,\dots, 8\}, \sigma =\downarrow \,,\\
        0\,, & \mbox{else}\,.
    \end{cases}
\end{align}
Here, it important to highlight that, although the constraints given by first two moments, cf. Eq.~\eqref{eq:first_moment_SMF} and \eqref{eq:second_moment_SMF}, are, by construction, exactly fulfilled, higher moments deviate heavily. More specifically, the $n$-th moment is of the order of $M^{(n-2)/2}$ for $n> 2$. Thus, a solution of this form is not useful within the context of the standard SMF approach. However, as the solution of the system of nonlinear equations given by Eqs.~\eqref{eq:deterministic_sampling_1} and \eqref{eq:deterministic_sampling_2} does admit multiple solutions, it is possible to find other solutions that minimize the deviations for higher moments. Further, this approach can be extended to include moments of higher order or systems with finite temperature or other spin configurations. In the context of approximations within the SMF approach and their applications, however, it is sufficient to consider the solution presented here.

\subsection{Stochastic polarization approximation (SPA)} \label{ss:SPA}
It is now possible to combine the SMF approach with the approximations in the fluctuation approach. On the one hand, this has the advantage that a treatment of two-particle approximations at the one-particle level becomes possible; on the other hand, the neglect of three-particle fluctuations leads to an inaccurate description of higher moments, which significantly influences the dynamics. In fact, it can be observed here that in the framework of two-particle approximations, all ensembles reproducing the first two moments are equivalent\cite{schroedter_cmp_22}. Equivalent in the sense that the dynamics of the system is the same. Therefore, it is no longer necessary to explicitly specify how the initial state was generated. The combination of SMF theory and QPA leads to the so-called stochastic polarization approximation (SPA)\cite{schroedter_cmp_22}. Here, the set of coupled equations is given by
\begin{gather}
    \mathrm{i}\hbar\partial_t G^<_{ij}(t)= \left[h^\mathrm{H},G^<\right]_{ij}(t)+\left[S+S^\dagger\right]_{ij}(t)+\overline{\left[ \Delta\Sigma^{\mathrm{H},\lambda},\Delta G^{\lambda} \right]}_{ij}(t)\,, \label{eq:EOM:G1-SPA}\\
    \mathrm{i}\hbar\partial_t\Delta G^\lambda_{ij}(t)=\left[h^\mathrm{HF},\Delta G^\lambda\right]_{ij}(t)+\left[ \Delta\Sigma^{\mathrm{HF},\lambda}, G^< \right]_{ij}(t)\,, \label{eq:EOM:single-particle_fluctuations_SPA}
\end{gather}
where $\Delta \Sigma^{\mathrm{HF},\lambda}$ denotes the random realization of the fluctuations Hartree--Fock selfenergy $\delta\hat{\Sigma}^\mathrm{HF}$.\\
Comparing the equations of the QPA, cf. Eqs.~\eqref
{eq:EOM:G1} and~\eqref{eq:EOM:L_QPA}, and SPA, cf. Eqs.~\eqref{eq:EOM:G1-SPA} and \eqref{eq:EOM:single-particle_fluctuations_SPA}, shows that the scaling of the CPU time is given by $\mathcal{O}(N_\mathrm{b}^6N_\mathrm{t})$ for QPA and by $\mathcal{O}(N_\mathrm{s}N_\mathrm{b}^4N_\mathrm{t})$ for SPA, where $N_\mathrm{b}$ denotes the number of considered basis states of the underlying Hilbert space, $N_\mathrm{t}$ the number of time steps and $N_\mathrm{s}$ the number of random realizations. Further, the scaling of the memory consumption is given by $\mathcal{O}(N_\mathrm{b}^4)$ for QPA and by $\mathcal{O}(N_\mathrm{s}N_\mathrm{b}^2)$ for SPA. This shows that the application of the SMF approach reduces the polynomial scaling with the number of basis states, but introduces a linear dependence on the number of random realizations. This, however, turns out to be a significant advantage of this approach. Within stochastic sampling, for example, the number of samples can be chosen constant so that the numerical scaling is given by  $\mathcal{O}(N_\mathrm{b}^4N_\mathrm{t})$ (CPU time) and $\mathcal{O}(N_\mathrm{b}^2)$ (memory) corresponding to the scaling of mean-field calculations. As the number of samples has to be chosen to be about $10^4$, this approach to the construction of the initial state is mostly advantageous for large systems. Given a system of electrons at zero temperature, deterministic sampling leads to $N_\mathrm{s}\sim N_\mathrm{p}N_\mathrm{h}$. Thus, the overall scaling using this approach is the of the QPA if $N_\mathrm{p}\approx N_\mathrm{h}$. For systems with $N_\mathrm{p}\ll N_\mathrm{h}$ or $N_\mathrm{h}\ll N_\mathrm{p}$, however, we have $N_\mathrm{s}\ll N_\mathrm{b}^2$, thus significantly reducing the computational scaling compared to the QPA.

\subsection{Multiple ensembles approach and SPA-ME} \label{ss:SPA-ME}
Although SMF theory allows to approximately solve operator equations and, thus, significantly reduces the computational effort, there are some shortcomings associated with replacing quantum-mechanical operators with an ensemble of realizations. Due to the semiclassical nature of the approach it is, for example, not possible to capture quantum effects like coherence. Hence, this approach is, by construction, restricted to weakly to moderately coupled systems. Moreover, as the random realizations commute whereas this is generally not the case for operators, it is not possible to compute any observable that depends on the specific ordering of the underlying operators, e.g., the retarded component of the density response function $\chi^\mathrm{R}$, cf. Eq.~\eqref{eq:definition:Chi}, where we have $\chi^{\mathrm{R},\mathrm{SMF}}\equiv 0$. This issue can be circumvented by instead replacing operators with noncommuting quantities. Within the framework of the approximations of the quantum fluctuations approach, this can be done using the so-called multiple ensembles (ME) approach\cite{schroedter_23}. Here, multiple ensembles are introduced, i.e., 
\begin{equation}
    \delta\hat{G}_{ij}\rightarrow \left( \Delta G^{(1),\lambda}_{ij},\Delta G^{(2),\lambda}_{ij}\right)\,,
\end{equation}
and products of operators are replaced according to their ordering, i.e., 
\begin{equation}
    \delta\hat{G}_{ij}\delta\hat{G}_{kl}\rightarrow \Delta G^{(1),\lambda}_{ij}\Delta G^{(2),\lambda}_{kl}\,.
\end{equation}
This, however, directly implies that this approach can only be applied within the framework of any approximation to the quantum fluctuations hierarchy where the considered equations are either linear or quadratic in the appearing single-particle fluctuations, i.e., it is not possible to apply this procedure to Eq.~\eqref{eq:EOM:quantum_single-particle_fluctuations}.\\
The two ensembles are then constructed according to 
\begin{gather}
    \overline{\Delta G^{(1),\lambda}_{ij}(t_0)}=\overline{\Delta G^{(2),\lambda}_{ij}(t_0)}=0\,,\label{eq:first_moment_ME}\\
    \overline{\Delta G^{(1),\lambda}_{ij}(t_0)\Delta G^{(2),\lambda}_{kl}(t_0)}=-\frac{1}{\hbar^2}\delta_{il}\delta_{jk}n_j(1\pm n_i)\,. \label{eq:second_moment_ME}
\end{gather}
Additionally, as the single-particle fluctuations operator obeys the symmetry $\delta\hat{G}_{ij}=-[\delta\hat{G}_{ji}]^\dagger$, the two ensembles have to obey an analogous symmetry relation given by
\begin{equation}
    \Delta G^{(1),\lambda}_{ij}=-\left[ \Delta G^{(2),\lambda}_{ji} \right]^*.
\end{equation}
Combining the ME approach and the SPA leads to the so-called SPA-ME. Here, the EOMs are given by
\begin{gather}
        \mathrm{i}\hbar\partial_t G^<_{ij}(t)= \left[h^\mathrm{H},G^<\right]_{ij}(t)+\left[S+S^\dagger\right]_{ij}(t)+\frac{1}{2}\left\{\overline{\left[ \Delta\Sigma^{\mathrm{H},(1),\lambda},\Delta G^{(2),\lambda} \right]}_{ij}(t)+\overline{\left[ \Delta\Sigma^{\mathrm{H},(2),\lambda},\Delta G^{(1),\lambda} \right]}_{ij}(t)\right\}\,, \label{eq:EOM:G1-SPA-ME}\\
    \mathrm{i}\hbar\partial_t\Delta G^{(m),\lambda}_{ij}(t)=\left[h^\mathrm{HF},\Delta G^{(m),\lambda}\right]_{ij}(t)+\left[ \Delta\Sigma^{\mathrm{HF},(m),\lambda}, G^< \right]_{ij}(t)\,. \label{eq:EOM:single-particle_fluctuations_SPA-ME}
\end{gather}
 Here, Eq.~\eqref{eq:EOM:G1-SPA-ME} is still expressed in a symmetrized form. This is  due to the breaking of exchange symmetries within the polarization approximation that may lead to instabilities in the numerical solution of the equations. However, these instabilities can be avoided by considering a symmetrized form of the EOM for the single-particle Green function. \\
 Numerical calculations of the SPA-ME have the same scaling as calculations within the SPA as the underlying equations, cf. Eqs.~\eqref{eq:EOM:G1-SPA}, \eqref{eq:EOM:single-particle_fluctuations_SPA}, \eqref{eq:EOM:G1-SPA-ME} and \eqref{eq:EOM:single-particle_fluctuations_SPA-ME}, do not differ in their complexity. Moreover, the same sampling methods can be used in both approaches. This is straightforward for the random creation of the initial state, but also optimized sampling algorithms can be easily adapted to the framework of the ME approach. However, within the SPA-ME, it is possible to meaningfully define quantities such as the retarded component of the density response function, i.e.,
 \begin{equation}
     \chi^{\mathrm{R},\mathrm{ME}}_{ij}(t_1,t_2)\coloneqq \mathrm{i}\hbar \Theta (t_1-t_2) \left[\overline{\Delta G^{(1),\lambda}_{ii}(t_1)\Delta G^{(2),\lambda}_{jj}(t_2)}-\overline{\Delta G^{(1),\lambda}_{jj}(t_2)\Delta G^{(2),\lambda}_{ii}(t_1)}\right]\,.
 \end{equation}
 In this sense, the SPA-ME constitutes an extension of the standard SPA while, at the same time, not increasing the numerical cost of simulations. Most importantly, this approach provides access to (spectral) two-particle observables, e.g., $\chi^\mathrm{R}$, while only having the structure of a set of mean-field equations. Additionally, it can be applied to describe systems in equilibrium as well as systems far from it.

\section{Numerical illustrations}\label{s:numerics}
The classical fluctuations approach discussed in Sec.~\ref{s:classical_fluctuations} was primarily used to derive kinetic equations for gases and plasmas, but not for direct numerical solutions for many-particle systems. At the same time, this approach has certain attractive features for numerical applications. In the following we will discuss possible computational applications, thereby focusing on the more general case of quantum systems. Most of the concepts can be applied in, similar form, also to classical systems. 

The quantum fluctuations approach discussed in this paper is completely general and applicable to any correlated quantum system. The details depend on the chosen single-particle basis $(\psi_i)_i$. Examples of interest would be atoms and molecules -- then one would use a basis of electronic orbitals. Another example are uniform systems such as the warm dense uniform electron gas, e.g.~[\onlinecite{dornheim_physrep_18}], electron-hole plasmas or dense quantum plasmas. A particular easy case to apply   the approach to are lattice systems, as those allow for simple numerical tests of the theory. 

\subsection{Lattice systems} \label{ss:lattice_systems}
Lattice models, such as the Fermi--Hubbard model or PPP model [\onlinecite{joost_phd_2022}], play an important role for the description of correlated electrons in condensed matter. Moreover, they can be used to model cold atoms in optical lattices. Another key factor is that an exact description, for example using exact diagonalization, is possible in special cases. Additionally, extensive data from NEGF and G1--G2 calculations are available to allow for benchmarks. \\
Within the Fermi--Hubbard model, the second quantization Hamiltonian, cf. Eq.~\eqref{eq:definition:Hamiltonian}, simplifies considerably: the single-particle contribution and the pair-interaction take the form
\begin{align}
    h_{ij} &\longrightarrow -J\delta_{\langle i,j\rangle} \,,\\
    w_{ijkl}^{\sigma_1\sigma_2\sigma_1'\sigma_2'} &\longrightarrow U\delta_{ij}\delta_{ik}\delta_{il}\delta_{\sigma_1\sigma_1'}\delta_{\sigma_2\sigma_2'}(1-\delta_{\sigma_1\sigma_2})\,,    \label{eq:Hubbard_interaction}
\end{align}
where $\delta_{\langle i,j\rangle}=1$, if the sites $i$ and $j$ are adjacent, and $\delta_{\langle i,j\rangle}=0$, if they are not. Further, $J$ denotes the hopping parameter describing hopping between neighboring sites and $U$ denotes the on-site interaction of two electrons on the same site. 
The Fermi--Hubbard Hamiltonian is then given by
\begin{equation}
    \hat{H}_\mathrm{FH}= -J \sum_{\langle i,j\rangle}\sum_{\sigma\in\{\uparrow,\downarrow\}} \hat{c}^\dagger_{i\sigma}\hat{c}_{j\sigma}+U\sum_i \hat{n}^\uparrow_i\hat{n}_i^\downarrow\,,
\end{equation}
where $\hat{n}^\sigma_i\coloneqq \hat{c}^\dagger_{i\sigma}\hat{c}_{i\sigma}$.\\
Within the SPA, the set of EOMs, cf. Eqs.~\eqref{eq:EOM:G1-SPA} and \eqref{eq:EOM:single-particle_fluctuations_SPA}, take the form 
\begin{gather}
    \mathrm{i}\hbar\partial_t G^{<,\sigma}_{ij}(t)= \left[h^{(1),\sigma}, G^{<,\sigma} \right]_{ij}(t)+\left[I+I^\dagger\right]^\sigma_{ij}(t)\,,\\
    \mathrm{i}\hbar\partial_t \Delta G^{\lambda,\sigma}_{ij}(t)=\left[h^{(1),\sigma},\Delta G^{\lambda, \sigma}\right]_{ij}(t)+\left[ \Delta\Sigma^{\lambda,\sigma},G^{<,\sigma} \right]_{ij}(t)\,,
\end{gather}
where, due to the form of the pair-interaction, cf. Eq.~\eqref{eq:Hubbard_interaction}, exchange contributions to the Hartree--Fock Hamiltonian and selfenergy vanish, i.e., we have
\begin{align}
    h^{(1),\sigma}_{ij}(t)&\coloneqq h^{\mathrm{H},\sigma}_{ij}(t)\equiv h^{\mathrm{HF},\sigma}_{ij}(t)= -J\delta_{\langle i,j\rangle }-\mathrm{i}\hbar U \delta_{ij} G^{<,\Bar{\sigma}}_{ii}(t)\,,\\
    \Delta \Sigma^{\lambda,\sigma}_{ij}(t)&\coloneqq\Delta \Sigma^{\mathrm{H},\lambda,\sigma}_{ij}(t)\equiv \Delta\Sigma^{\mathrm{HF},\lambda,\sigma}_{ij}(t)= -\mathrm{i}\hbar U\delta_{ij} \Delta G^{\lambda,\Bar{\sigma}}\,,
\end{align}
where $\sigma=\uparrow (\downarrow)$ implies $\Bar{\sigma}=\downarrow (\uparrow)$. Further, the collision term is given by
\begin{equation}
    I^\sigma_{ij}(t)= -\mathrm{i}\hbar U\overline{\Delta G^{\lambda,\Bar{\sigma}}_{ii}(t)\Delta G^{\lambda,\sigma}_{ij}(t)}\,.
\end{equation}
Within the Hubbard model, the symmetrization contributions, cf. Eq.~\eqref{eq:definition:S-term}, vanish. Further, the initial state of the system is (in the basis of natural orbitals) chosen such that 
\begin{align}
    G^{<,\sigma}_{ij}(t_0) &= -\frac{1}{\mathrm{i}\hbar} \delta_{ij}n_i^\sigma,\\
    \overline{\Delta G^{\lambda,\sigma}_{ij}(t_0)} &= 0\,,\\
    \overline{\Delta G^{\lambda,\sigma}_{ij}(t_0)\Delta G^{\lambda,\sigma'}_{kl}(t_0)} &= -\frac{1}{2\hbar^2}\delta_{il}\delta_{jk}\delta_{\sigma\sigma'} \left(1-\delta_{n_i^\sigma n_j^{\sigma'}}\right)\,.
\end{align}
For the SPA-ME, we instead consider the corresponding expressions to Eqs.~\eqref{eq:first_moment_ME} and \eqref{eq:second_moment_ME}.
Based on the system's configuration, a transformation is necessary from the basis of natural orbitals to the Hubbard basis. Furthermore, a nontrivial ground state is generated using the so-called ``adiabatic switching method'' [\onlinecite{schluenzen_prb16}]. Here, the on-site interaction $U$ is replaced by a time-dependent interaction, $U(t)$, which is chosen such that $U(t)$ increases monotonically and sufficiently slowly and satisfies $U(t_s) =0 $ and $U(t)= U$ for $t\geq t_0$, where $t_s$ denotes the time at which the simulations start with an uncorrelated ideal ground state. \\
In the following, we consider one-dimensional Hubbard chains with periodic conditions (PBC). In particular, we consider systems in the ground state as well as system after an excitation given by a so-called ''confinment quench'' \cite{schluenzen_prb16}. Here, the particles are confined to a set of connected sites so that these are fully occupied (given an even number of particles) while the remaining are empty. At $t_0$, the confinement is lifted, leading to a diffusion-type process in the system. \\
Here, we extend the results presented in Refs.~\onlinecite{schroedter_cmp_22,schroedter_23,schroedter_pssb23,bonitz_pssb23} and first discuss different sampling methods in Sec.~\ref{ss:comparison_sampling} In particular, we consider results for the density on the first site given by
\begin{equation}
    n_1(t)\coloneqq -\mathrm{i}\hbar\left[ G^{<,\uparrow}_{11}(t)+G^{<,\downarrow}_{11}(t)\right]\,,
\end{equation}
following a confinement quench and compare SPA-data using different sampling methods (deterministic sampling and stochastic sampling with Gaussian and four-point distributions) to results from the QPA for systems of different size and with varying on-site interaction strength. This is followed by a comparison of the SPA to the nonequilibrium $GW$ approximation within the G1--G2 scheme and the RPA for a system in the ground state in Sec.~\ref{ss:comparison_approximations}. More specifically, we again consider nonequilibrium systems following a confinement quench and discuss results for the density, cf. Sec.~\ref{sss:neq_density} and, furthermore, test the SPA-ME for systems in the ground state, cf. Sec~\ref{sss:dynamic_structure_factor}, and nonequilibrium systems, cf. Sec.~\ref{sss:neq_density_response}. For this, we consider the density response function, cf. Eq.~\eqref{eq:definition:Chi} [we additionally sum over all spin configurations, i.e., $\chi^\mathrm{R}_{ij}(t,t')=\sum_{\sigma\sigma'}\chi^{\mathrm{R},\sigma\sigma'}_{ij}(t,t')$] and the (ground-state) dynamic structure factor given by
\begin{eqnarray}
    S(q,\omega) = -4\mathrm{Im}\left[ \chi^\mathrm{R}(q,\omega) \right]\,,
\end{eqnarray}
where the Fourier-transformed density response function $\chi^\mathrm{R}(q,\omega)$ follows from identifying the lattice site $i$ with the position $x_i\coloneqq ia_0$, where $a_0$ denotes the characteristic distance between two lattice sites, meaning the relative position is given by $r_{ij}\coloneqq (i-j)a_0$. Additionally, as we consider Fourier transforms over a finite time length, we introduce a additional factor of $e^{-\eta t}$ to mitigate some of the effects of not integrating over the real numbers. For information about the implementation of the (two-time) QPA and $GW$ approximation we refer to Ref.~\onlinecite{schroedter_pssb23} and for the G1--G2 scheme we refer to   Refs.~\onlinecite{schluenzen_prl_20,joost_prb_20}.
\subsection{Comparison of sampling methods} \label{ss:comparison_sampling}
\begin{figure}
    \centering
    \includegraphics[width=1.0\textwidth]{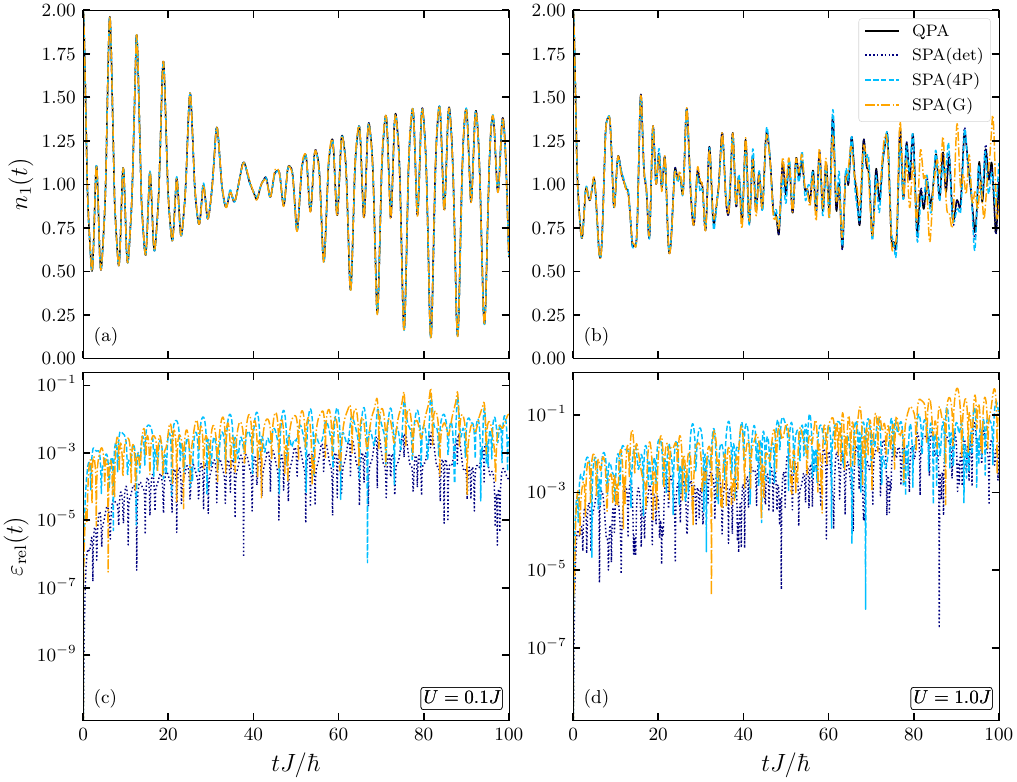}
    \caption{Comparison of the density on the first site for different sampling methods within the SPA to the QPA for a half-filled six-site chain at $U=0.1J$ and $U=1.0J$ following a confinement quench with the left-most 3 sites fully occupied. The uncorrelated initial states for the SPA calculations were generated using deterministic (''$\mathrm{det}$'') and stochastic sampling. For the latter, complex four-point (''$\mathrm{4P}$'') and Gaussian (''$\mathrm{G}$'') distributions with $10^4$ random realizations for each spin configuration were used. Deterministic sampling used 36 samples for each spin component. (a) and (b) show the density on the first site for $U=0.1J$ and $U=1.0J$, respectively. (c) and (d) display the relative deviation of the sampling approaches to the results from the QPA, which is given by $\varepsilon_\mathrm{rel}(t)\coloneqq |1-n_1^\mathrm{SPA}(t)/n_1^\mathrm{QPA}(t)|$.}
    \label{fig:comparison_sampling_ns6}
\end{figure}

\begin{figure}
    \centering
    \includegraphics[width=1.0\textwidth]{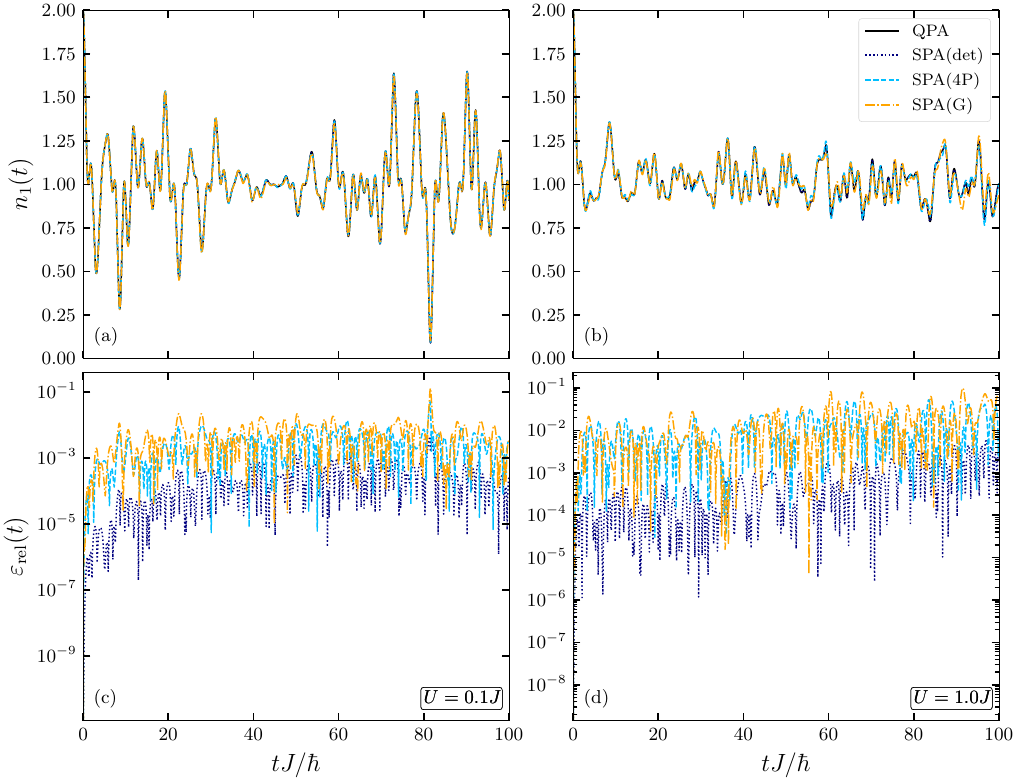}
    \caption{Same as Fig.~\ref{fig:comparison_sampling_ns6}, but for a half-filled ten-site chain.}
    \label{fig:comparison_sampling_ns10}
\end{figure}

As we discussed in Sec.~\ref{ss:sampling}, the standard SMF approach relies heavily on the appropriate construction of the quantum initial state. However, it is impossible to exactly solve this problem within the framework of the standard approach and, thus, one of the main tasks becomes finding the best approximate solution. It is therefore of great interest to analyze the effects of combining SMF theory with the approximations within the quantum fluctuations approach to see whether it is possible to mitigate some of the effects of semi-classically generating an initial state and whether it is possible to adequately solve operator equations, such as Eq.~\eqref{eq:EOM:quantum_single-particle_fluctuations_QPA}, using the SMF approach.  In Fig.~\ref{fig:comparison_sampling_ns6}, we first consider a half-filled system with six sites following a confinement quench. For $U=0.1J$, we see that there are only minor deviations of the SPA to the QPA for all sampling methods. While the relative deviation for stochastic sampling remains within $10^{-3}$ to $10^{-1}$ in panel (c), the relative deviation for deterministic sampling is below $10^{-3}$. Both sampling approaches, however, do not show a significant increase of deviations for longer times. For stronger coupling, i.e., $U=1.0J$, we find that deviations between the QPA and the SPA increase more strongly compared to $U=0.1J$ and visible differences can be seen in (b) for the stochastic sampling approaches when directly considering the density $n_1$. Relative deviations, shown in panel (d), increase in this case up to $10^{-1}$, while, for deterministic sampling, deviations remain smaller ($\sim 10^{-2}$). \\
For a larger system with ten sites, Fig.~\ref{fig:comparison_sampling_ns10} shows that deviations are smaller for $U=0.1J$ and $U=1.0J$ compared to the six-site chain at the respective coupling strengths. At $U=0.1J$, we see in panel (c) that the results from the SPA using stochastic sampling are $\sim 10^{-3}$ for almost all times, whereas the results obtained using deterministic sampling are $\sim 10^{-4}$ in that range. For stronger on-site interaction strength, we see in panels (b) and (d) that there are again more significant deviations compared to the former case, however, with respect to the smaller system, we find that deviations for stochastic sampling are $\sim 10^{-2}$ while they are only $\lesssim 10^{-3}$ for deterministic sampling.\\
These results imply that the SPA and QPA are equivalent under the condition that the first two moments of the initial state are exactly reproduced. In particular, this shows that all probability distributions with the correct first two moments are equivalent (in the sense of leading to the same dynamics of the system). Moreover, the dynamics of a system given randomly generated realizations of the initial state converges to the dynamics of a system with the exactly reproduced initial state (within the setting of the QPA). This convergence, however, depends on the system size and the coupling strength, i.e., stronger coupling makes the dynamics more sensitive to deviations from the correct initial state, whereas the rate of convergence improves for larger systems. This is further investigated in Ref.~\onlinecite{schroedter_cmp_22} for larger systems. For this reason, it is possible to choose a sufficiently large number of samples for the ensemble, which is independent of the system size. In this case, SPA calculations have the same numerical scaling as simple mean-field calculations while still corresponding to the QPA, thus making this approach perfectly suited for the description of large quantum systems that are weakly coupled.\\
Furthermore, these results also illustrate that, although the standard SMF approach is only able to approximately solve the operator equation given by Eq.~\eqref{eq:EOM:single-particle_NEGF_operator}, i.e., the full $N$-body problem, the SMF approach combined with the quantum polarization approximation allows for an (almost) exact solution of the operator equation given by Eq.~\eqref{eq:EOM:quantum_single-particle_fluctuations_QPA}. Although applying the QPA to the fluctuations hierarchy itself constitutes an attempt to approximately solve the $N$-body problem, this application of SMF theory can be considered more advantageous compared to the standard approach because it eliminates the dependence on reproducing all moments of the initial state while retaining the favorable aspects of SMF calculations in terms of numerical scaling.
\subsection{Comparison of approximations} \label{ss:comparison_approximations}
As was mentioned in Sec.~\ref{sss:QPA}, the QPA is closely related to the $GW$ approximation within the G1--G2 scheme and is, in particular, equivalent to the $GW$ approximation with additional exchange contributions in the weak coupling limit. In the following, we will analyze this relation in more detail by comparing results obtained from the SPA using deterministic sampling to $GW$ results within the G1--G2 scheme. First, we consider the density dynamics following a confinement quench. Then, we compare ground-state results for the dynamic structure for the SPA-ME and the RPA and nonequilibrium results for the density response function using the QPA, SPA-ME and $GW$ approximation.
\subsubsection{Nonequilibrium density following a confinement quench} \label{sss:neq_density}
\begin{figure}[h]
\centering
\includegraphics[width=1.00\columnwidth]{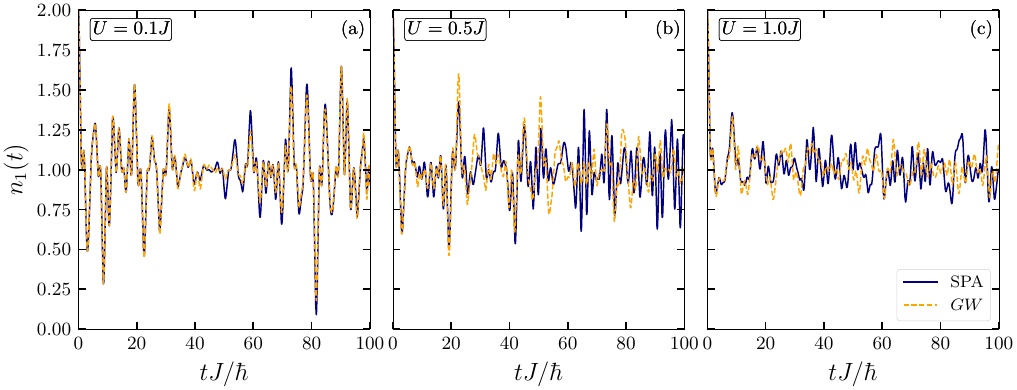}
\caption{Evolution of the density on the first site of a half-filled ten-site Hubbard chain at $U=0.1J$ (a), $U=0.5J$ (b) and $U=1.0J$ (c) following a confinement quench. In the uncorrelated initial state, the leftmost five sites are fully occupied whereas the remaining are empty. The SPA using deterministic sampling is compared to the $GW$ approximation within the G1--G2 scheme.}
\label{fig:comparison_density}
\end{figure}
Figure~\ref{fig:comparison_density} shows the density dynamics of a half-filled chain with ten sites following a confinement quench at $U/J= 0.1, 0.5,1.0$. For $U=0.1J$, we see in panel (a) that there is very good agreement between the SPA and $GW$ approximation. We see that for times $t\lesssim 40\hbar/J$, there are almost no visible deviations between the two approximations. Only for later times, differences between the two approximations become more apparent. However, it is important to note that these differences are mainly limited to the amplitudes of the oscillations, while their frequencies still agree very well. At $U=0.5J$, shown in panel (b), the SPA and $GW$ approximation agree very well for times $t\lesssim 25\hbar/J$. Then, for times until $t\sim 60\hbar/J$, there is still good agreement between the two approximations. However, after this point, the agreement is mostly of qualitative nature. Here, we see that, similar to the case of $U=0.1J$, the SPA tends to overestimate the amplitude of the oscillations compared to the $GW$ approximation. Additionally, we observe that the frequencies are also significantly larger for the SPA than for the $GW$ approximation. The same behavior can be seen in panel (c) for the system at $U=1.0J$. Here, we find that there is very good agreement between the SPA and the $GW$ approximation up to a time $t\sim 20\hbar/J$. After this point, there is only qualitative agreement between the two approximations. However, as both approximations are based on the assumption of weak coupling, neither of them are applicable for a nonequilibrium system at this coupling strength. This is also illustrated in Ref.~\onlinecite{schroedter_cmp_22} for a system with eight sites, where it is possible to exactly solve the equations of motion using exact diagonalization.  There it can be seen that both fail to accurately reproduce the density dynamics following a confinement quench.\\
The extended analysis of the relation between the SPA and $GW$ approximation, done here and in Ref.~\onlinecite{schroedter_cmp_22}, shows that the two approximations can be considered equivalent in the weak coupling limit. However, a key difference between the two approximations is given by their possible implementation and numerical scaling. The $GW$ approximation has, within the G1--G2 scheme\cite{joost_prb_20}, the same numerical scaling as the QPA, i.e., $\mathcal{O}(N_\mathrm{b}^6N_\mathrm{t})$ for the CPU time and $\mathcal{O}(N_\mathrm{b}^4)$ for the memory consumption (for a general basis). On the other hand, the SPA has a scaling given by $\mathcal{O}(N_\mathrm{s}N_\mathrm{b}^4N_\mathrm{t})$ and $\mathcal{O}(N_\mathrm{s}N_\mathrm{b}^2$) for CPU time and memory consumption, respectively. As the number of samples $N_\mathrm{s}$ can be chosen constant, the SPA effectively allows for calculations at the level of the $GW$ approximation while having the numerical scaling of simple mean-field equations.
\subsubsection{Ground-state dynamic structure factor} \label{sss:dynamic_structure_factor}
\begin{figure}
    \centering
    \includegraphics[width=0.85\textwidth]{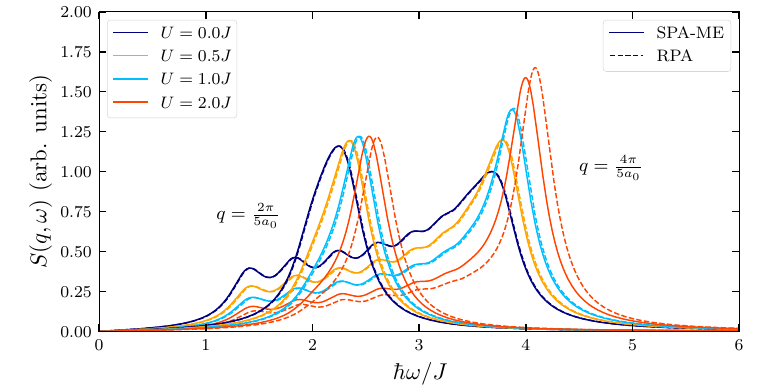}
    \caption{Dynamic structure factor of a half-filled fifty-site chain for different on-site interaction strengths at $q=2\pi/(5a_0)$ and $q=4\pi/(5a_0)$ using SPA-ME and RPA. A damping constant of $\eta = 0.2 J/\hbar$ was used. %with a damping constant of $\eta = 0.02 J/\hbar$.
    }
    \label{fig:dynamic_structure_factor}
\end{figure}
Despite the significant advantages the SPA provides, it still suffers from a defect inherent to SMF theory: the inability to compute two-particle observables depending on the ordering of the underlying operators. The proposed solution to this problem in the form of the multiple ensembles approach has the advantage that it preserves all the advantages of the SPA. However, from a theoretical point of view, it is not obvious why this approach should lead to any meaningful results. To demonstrate the validity of this approach and to further extend the results presented in Ref.~\onlinecite{schroedter_pssb23}, we first consider ground-state results for the dynamic structure factor of a half-filled fifty-site chain for the SPA-ME and RPA. Here we use a damping constant of $\eta = 0.2J/\hbar$ for the exponential damping factor $e^{-\eta t}$ multiplied with the density response function $\chi^\mathrm{R}(t)$ for the SPA-ME. The same damping constant is use for the RPA expression for dynamic structure factor. Figure~\ref{fig:dynamic_structure_factor} shows the ground-state dynamic structure factor for a half-filled chain with fifty sites at different coupling strengths. Here, we see that the SPA-ME shows very good agreement with the RPA for weak coupling. Only for $U=2.0J$ do deviations become visible. For $q= 2\pi /(5a_0)$, we see that the main peak is located at $\omega \approx 2.5 J/\hbar$ with $S\approx 1.2 \mathrm{a.u.}$ for the SPA-ME whereas it is located at $\omega \approx 2.6 J/\hbar$ for the RPA also with $S\approx 1.2\mathrm{a.u.}$. Similarly, we find for $q= 4\pi/(5a_0)$ that the main peak for the SPA-ME is located at $\omega \approx 4.0J/\hbar$ with $S\approx 1.6\mathrm{a.u.}$ while it is shifted to higher frequencies for the RPA ($\omega \approx 4.3 J/\hbar$, $S\approx 1.7\mathrm{a.u.}$). Overall, we find that the main peaks shift for both wave numbers $q$ to higher frequencies with increasing coupling strength. For $q=4\pi/(5a_0)$, we see that the main peak is located at $\omega \approx 3.9 J/\hbar$ with $S\approx 1.0\mathrm{a.u.}$ for the noninteracting system and at $\omega \approx 4 J/\hbar$ with $S\approx 1.2\mathrm{a.u.}$ and $\omega \approx 4J/\hbar$ with $S\approx 1.4\mathrm{a.u.}$. Further, we find for $q=2\pi/(5a_0)$ that the main peaks are located at $\omega \approx 2.2 J/\hbar$ with $S\approx 1.1 \mathrm{a.u.}$ for the noninteracting system at $\omega \approx 2.35 J/\hbar$ for $U=0.5J$ and $\omega \approx 2.4 J/\hbar$ for $U=1.0J$ both with $S\approx 1.2\mathrm{a.u.}$. \\
Due to the finite size of the system, instead of a single peak for each wave number like for the infinite chain, the results for the dynamic structure factor show multiple peaks and a superposition of these leads to the main peaks. We see that for $q=2\pi/(5a_0)$ these peaks are positioned so that only a single peak is visible whereas for $q=4\pi/(5a_0)$ these peaks are clearly distinguishable. Further, it is shown that for increasing coupling strength these peaks become less pronounced, i.e., for the noninteracting system, we find that the peaks have a height of $S\approx 0.5\mathrm{a.u.}$, whereas for $U=2.0J$, the peaks are located at $S\approx 0.25 \mathrm{a.u.}$.\\
An important factor that leads to the visible differences between the SPA-ME and RPA results for moderate coupling is due to the different levels of selfconsistency of both approximations. While the SPA-ME considers correlated Green functions, ideal Green functions are considered for the RPA. The excellent agreement between the two approximations for weak coupling, however, shows that the multiple ensembles approach allows for the calculation of spectral two-particle observables like response functions and their dynamic structure factors while preserving the equivalence of the QPA and the $GW$ approximation/RPA. A key difference between the two methods, however, is that the SPA-ME is not restricted to the ground state and can also be applied to any nonequilibrium scenario.
\subsubsection{Nonequilibrium density response following a confinement quench} \label{sss:neq_density_response}
\begin{figure}
    \centering
    \includegraphics{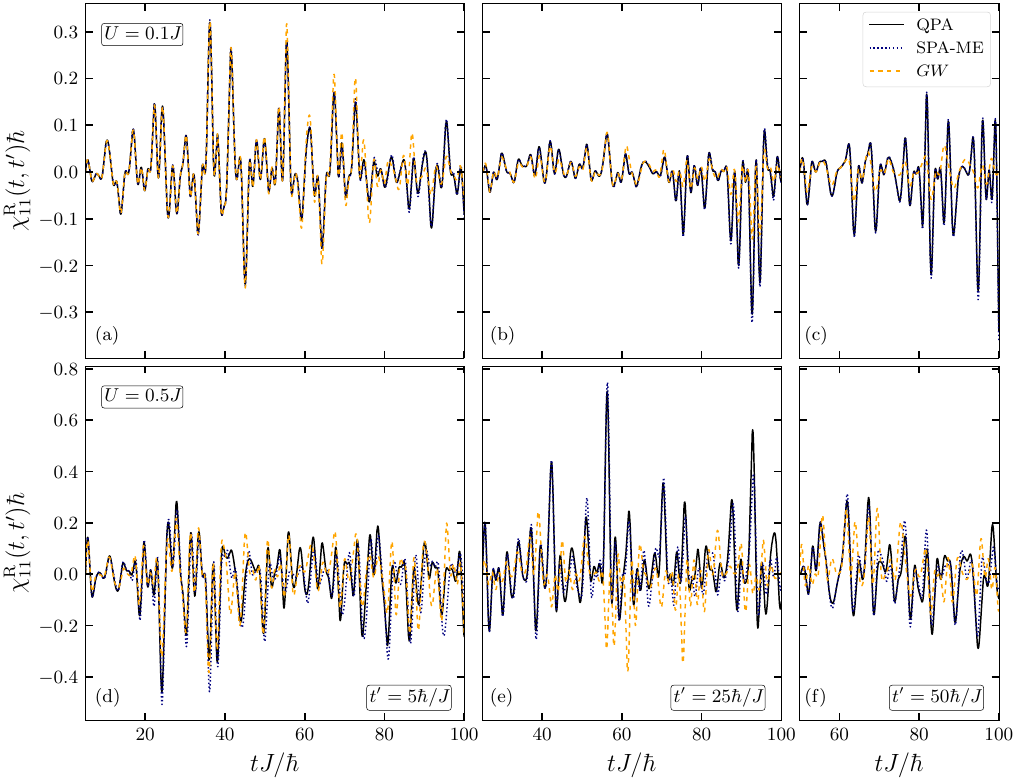}
    \caption{Retarded component of the density response function for a half-filled ten-site Hubbard chain at $U/J=0.1, 0.5$ (rows), following a confinement quench, for fixed $t'J/\hbar =  5, 25, 50$ (columns). The initial state was uncorrelated and with the five left-most sites fully occupied. The results for the QPA and SPA-ME are compared to data from the $GW$ approximation.}
    \label{fig:chi}
\end{figure}
One of the most important aspects of the SPA is its favorable numerical scaling. Extending the SPA by means of the multiple ensembles approach preserves this aspect and allows for the calculation of nonequilibrium spectral two-particle observables from single-time mean-field equations. To demonstrate that the applicability of the SPA-ME extends to nonequilibrium  scenarios and is not restricted to the ground state, we again consider a half-filled ten-site chain following a confinement quench and compare the SPA-ME to the two-time QPA and $GW$ approximation. Figure~\ref{fig:chi} shows the density response function for the chain at $U=0.1J$ and $U=0.5J$. For $U=0.1J$, we see excellent agreement of the SPA-ME and the QPA. However, for later times, i.e., $t\gtrsim 80 \hbar/J$ for all $t'$, the amplitudes of the oscillations are slightly larger for the SPA compared to the QPA. Nonetheless, the frequencies perfectly agree. This is also seen when considering the $GW$ approximation compared to the polarization approximations. While for $t'= 5\hbar/J$, shown in panel (a), the amplitudes of the oscillations are larger for times $60\hbar/J \lesssim t \lesssim 85 \hbar/J$, for all other times $t$ and $t'$, the amplitudes for the $GW$ are generally smaller than for the polarization approximations. Nonetheless, we seen in panels (a)-(c) that there is very good agreement between the approximations for almost all times. At stronger coupling, we see that deviations between the three approximations increase. While for $t'=5\hbar/J$, shown in panel (d), the agreement between the polarization approximations is very good for all times, we see deviations arise compared to the $GW$ approximation for times $t\gtrsim 40\hbar/J$. However, we again find that the frequencies agree very well and only deviations for the amplitudes of the oscillations are noticeable. For $t'=25\hbar/J$, shown in panel (e), there is mostly qualitative agreement of the $GW$ approximation and the polarization approximations. In panel (f), we see for $t'=50\hbar/J$ that the agreement improves compared to the results shown in panel (e).\\
These results illustrate the capability of the SPA-ME to calculate dynamic response functions for systems far from equilibrium. In particular, they show that the SPA-ME and QPA can be considered equivalent. Further, it is important to highlight that the deviations between the SPA-ME and QPA that are visible for stronger coupling are only observable for smaller systems, i.e. they are mainly caused by finite size effects. For a system with thirty sites following a confinement quench, the SPA-ME and QPA perfectly agree for stronger coupling~[\onlinecite{schroedter_pssb23}]. Moreover, these results show that the close correspondence between the QPA and the $GW$ also extends to the two-time case. Here, however, spectral two-particle observables, like the density response function, are much more sensitive to correlations in the system (we have $\chi\sim L$), so that deviations are more pronounced between the approximations compared to observables that depend on the time-diagonal single-particle Green function, e.g., the density $n$. Overall, we find that the SPA-ME provides easy and cost-effective access to spectral two-particle observables for nonequilibrium systems while retaining its close relation to the nonequilibrium $GW$ approximation.

\section{Conclusions and outlook}\label{s:discussion}
In this article which was devoted to the 100th birthday of Yuri L. Klimontovich, we gave a brief summary of his classical fluctuations approach which is based on the microscopic phase space density $N(x,t)$ and its fluctuations. There is a one to one correspondence between correlation functions of fluctuations and pair and triple correlations, i.e. between the hierarchy of equations for fluctuations and the BBGKY-hierarchy. This allows one to identify the approximations of the fluctuations approach that lead to other known collision integrals, e.g.~[\onlinecite{kolberg_phys-rep_18}].
As an important application of the fluctuations approach we demonstrated how the derivation of collision integrals of kinetic equations can be straightforwardly performed which was illustrated on the example of the  Balescu-Lenard integral, in Sec.~\ref{sss:ble}. We also discussed limitations and extensions of the fluctuations approach. In particular, we discussed the quantum generalization of the Balescu-Lenard equation and how to extend it to short-time processes, cf. Sec.~\ref{sss:ble-short}.

The main part of this article was devoted to the extension of the fluctuations approach to quantum systems. In Sec.~\ref{s:quantum_fluctuations} we introduced the theoretical basis that builds on fluctuations of nonequilibrium Green functions operators, $\delta \hat G$, i.e. field operator products, and introduced the correlation functions of fluctuations. The central role is being played by the exchange correlation function $L$, i.e., the exchange and correlation contributions to the two-particle NEGF. As in the classical case, this function contains induced fluctuations (due to correlations, $\mathcal{G}$) and spontaneous fluctuations, $L^0$, which in quantum systems have the form of particle-hole pair excitations and take over the role of the classical source fluctuation, $\delta N^\mathrm{S}$. We then turned to the equations of motion for the fluctuations and discussed important approximations for decoupling of the resulting hierarchy. The most important approximation was shown to be the quantum polarization approximation (QPA) -- coupled equations for the single-particle Green function $G$, Eq.~\eqref{eq:EOM:G1} and the two-particle fluctuation $L$, Eq.~\eqref{eq:EOM:L_QPA}. On the one hand, it constitutes the quantum generalization of the classical polarization approximation (that leads to the Balescu-Lenard equation, Sec.~\ref{sss:ble}) and is equivalent to an important approximation of quantum many-body physics -- the $GW$ approximation. On the other hand, the QPA equation for $L$ has the interesting property that it can be re-expressed via an equation of motion for the single-particle fluctuation, $\delta \hat G$, Eq.~\eqref{eq:EOM:quantum_single-particle_fluctuations_QPA}. The existence of this equation was the basis for eliminating two-particle quantities from the numerical propagation scheme the memory consumption of which constitutes the main bottleneck in the G1--G2 scheme [\onlinecite{bonitz_pssb23}]. Instead, we developed a semiclassical stochastic approach to sample random realizations of the single-particle quantity, $\delta \hat G \to \Delta G^{\lambda}$, which lead to the stochastic polarization approximation (SPA).

The ideas of this stochastic approach were outlined in Sec.~\ref{s:stochastic_approach}. There we discussed the properties of the associated random process, its moments and various concepts how to sample realizations of non-commuting operators via the extension of the scheme to two independent ensembles (``multiple ensembles approach'', ME, which was first introduced in Ref.~[\onlinecite{schroedter_23}]). We compared different sampling schemes and demonstrated that they agree very well with each other if they correctly reproduce the first two moments. An important application of the ME approach was the computation of the dynamic (frequency dependent) density response function, $\chi^R(t,t')$, and the dynamic structure factor $S(q,\omega)$,  which requires the computation of two-time correlation functions, involving products of $\delta \hat G$ taken at different times. These are quantities that are not available in the time local G1--G2 scheme. In contrast, the quantum fluctuations approach allows for a computation of these and similar observables in equilibrium but also in nonequilibrium, where the time dependence is not only given by the difference $t_1-t_2$, but involves also the center of mass time that reflects the dynamics of single-particle properties. Finally, the concept of the quantum fluctuations approach and its stochastic realization were numerically illustrated for lattice models, in Sec.~\ref{s:numerics}. As a generic nonequilibrium excitation scenario we considered a confinement quench to which the system (a Hubbard chain with $10 \dots 50$ lattice sites) responds with spatial diffusion that is strongly affected by the interaction strength, e.g.~[\onlinecite{schluenzen_prb16,schluenzen_prb17}].
The tests demonstrated that the quantum fluctuations approach and its stochastic implementation are indeed equivalent and also agree with the $GW$ approximation within the G1--G2 scheme within their applicability range, i.e. for weak coupling. This provides strong support for the validity of the present quantum fluctuations approach. 

Due to its advantageous scaling with respect to CPU time and memory, we expect that our quantum fluctuations approach can be efficiently applied to large lattice systems in nonequilibrium that contain hundreds of sites. Moreover, a promising application is to spatially uniform systems for which two-particle quantities require very large computer memory which currently limits the G1--G2 scheme to one-dimensional models [\onlinecite{makait_cpp_23,bonitz_pssb23}]. This includes the uniform electron gas, electron-hole plasmas and dense quantum plasmas. Further, it will be of interest to compute additional quantities such as spin response functions or the momentum distribution and to compare to benchmarks, such as quantum Monte Carlo results [\onlinecite{hunger_pre_21}]. An important question will be to find approximations that go beyond the applicability limits of the QPA and allow one to access strong coupling situations. Finally, it will be interesting to compare the present quantum fluctuations approach to other stochastic schemes such as the truncated Wigner approximation [\onlinecite{polkovnikov_ap_10}] or 
 stochastic density functional theory (sDFT), e.g.~[\onlinecite{baer_prl_13}]. 

\section*{Acknowledgments}
We acknowledge fruitful discussions with J.-P. Joost, B.J. Wurst and C. Makait. This work has been supported by the Deutsche Forschungsgemeinschaft via grant BO1366/16. 

%\bibliography{paper.bib,mb-ref.bib,Literatur1}

%% \begin{thebibliography}
%% \bibitem[a] bc.
%% \end{thebibliography}

\providecommand{\url}[1]{\texttt{#1}}
\providecommand{\urlprefix}{}
\providecommand{\foreignlanguage}[2]{#2}
\providecommand{\Capitalize}[1]{\uppercase{#1}}
\providecommand{\capitalize}[1]{\expandafter\Capitalize#1}
\providecommand{\bibliographycite}[1]{\cite{#1}}
\providecommand{\bbland}{and}
\providecommand{\bblchap}{chap.}
\providecommand{\bblchapter}{chapter}
\providecommand{\bbletal}{et~al.}
\providecommand{\bbleditors}{editors}
\providecommand{\bbleds}{eds: }
\providecommand{\bbleditor}{editor}
\providecommand{\bbled}{ed.}
\providecommand{\bbledition}{edition}
\providecommand{\bbledn}{ed.}
\providecommand{\bbleidp}{page}
\providecommand{\bbleidpp}{pages}
\providecommand{\bblerratum}{erratum}
\providecommand{\bblin}{in}
\providecommand{\bblmthesis}{Master's thesis}
\providecommand{\bblno}{no.}
\providecommand{\bblnumber}{number}
\providecommand{\bblof}{of}
\providecommand{\bblpage}{page}
\providecommand{\bblpages}{pages}
\providecommand{\bblp}{p}
\providecommand{\bblphdthesis}{Ph.D. thesis}
\providecommand{\bblpp}{pp}
\providecommand{\bbltechrep}{}
\providecommand{\bbltechreport}{Technical Report}
\providecommand{\bblvolume}{volume}
\providecommand{\bblvol}{Vol.}
\providecommand{\bbljan}{January}
\providecommand{\bblfeb}{February}
\providecommand{\bblmar}{March}
\providecommand{\bblapr}{April}
\providecommand{\bblmay}{May}
\providecommand{\bbljun}{June}
\providecommand{\bbljul}{July}
\providecommand{\bblaug}{August}
\providecommand{\bblsep}{September}
\providecommand{\bbloct}{October}
\providecommand{\bblnov}{November}
\providecommand{\bbldec}{December}
\providecommand{\bblfirst}{First}
\providecommand{\bblfirsto}{1st}
\providecommand{\bblsecond}{Second}
\providecommand{\bblsecondo}{2nd}
\providecommand{\bblthird}{Third}
\providecommand{\bblthirdo}{3rd}
\providecommand{\bblfourth}{Fourth}
\providecommand{\bblfourtho}{4th}
\providecommand{\bblfifth}{Fifth}
\providecommand{\bblfiftho}{5th}
\providecommand{\bblst}{st}
\providecommand{\bblnd}{nd}
\providecommand{\bblrd}{rd}
\providecommand{\bblth}{th}

%------------------------------------------------------

%\appendix

\end{document}